\documentclass[journal=mamobx, manuscript=article]{achemso}

\usepackage{eurosym}
\usepackage{enumitem}
\usepackage{tabularx}
\usepackage{multirow}
\usepackage{graphics}
\usepackage{epsfig}
\usepackage{multirow}
\usepackage{amssymb}
\usepackage{amsmath}
\usepackage{leftidx}
\usepackage{epsfig}
\usepackage{color,soul}
\usepackage{array}
\usepackage{array, makecell}
\usepackage[explicit]{titlesec}

\title{Structure and Dynamics of Hybrid Colloid-Polyelectrolyte Coacervates: Insights from Molecular Simulations}

\author{Boyuan Yu}
\affiliation{Pritzker School of Molecular Engineering, University of Chicago, Chicago, Illinois 60637, United States}

\author{Heyi Liang}
\affiliation{Pritzker School of Molecular Engineering, University of Chicago, Chicago, Illinois 60637, United States}

\author{Paul F. Nealey}
\affiliation{Pritzker School of Molecular Engineering, University of Chicago, Chicago, Illinois 60637, United States}
\alsoaffiliation{Center for Molecular Engineering, Argonne National Laboratory, Lemont, Illinois 60439, United States}

\author{Matthew Tirrell}
\affiliation{Pritzker School of Molecular Engineering, University of Chicago, Chicago, Illinois 60637, United States}
\alsoaffiliation{Center for Molecular Engineering, Argonne National Laboratory, Lemont, Illinois 60439, United States}

\author{Artem~M.~Rumyantsev}
\affiliation{Pritzker School of Molecular Engineering, University of Chicago, Chicago, Illinois 60637, United States}
\alsoaffiliation{Department of Chemical and Biomolecular Engineering, North Carolina State University, Raleigh, North Carolina 27695-7905, United States}

\author{Juan~J.~de~Pablo}
\affiliation{Pritzker School of Molecular Engineering, University of Chicago, Chicago, Illinois 60637, United States}
\alsoaffiliation{Center for Molecular Engineering, Argonne National Laboratory, Lemont, Illinois 60439, United States}
\email{depablo@uchicago.edu}


\DeclareUnicodeCharacter{2212}{-}
\begin{document}

\begin{tocentry}
\includegraphics[width = 3.25in]{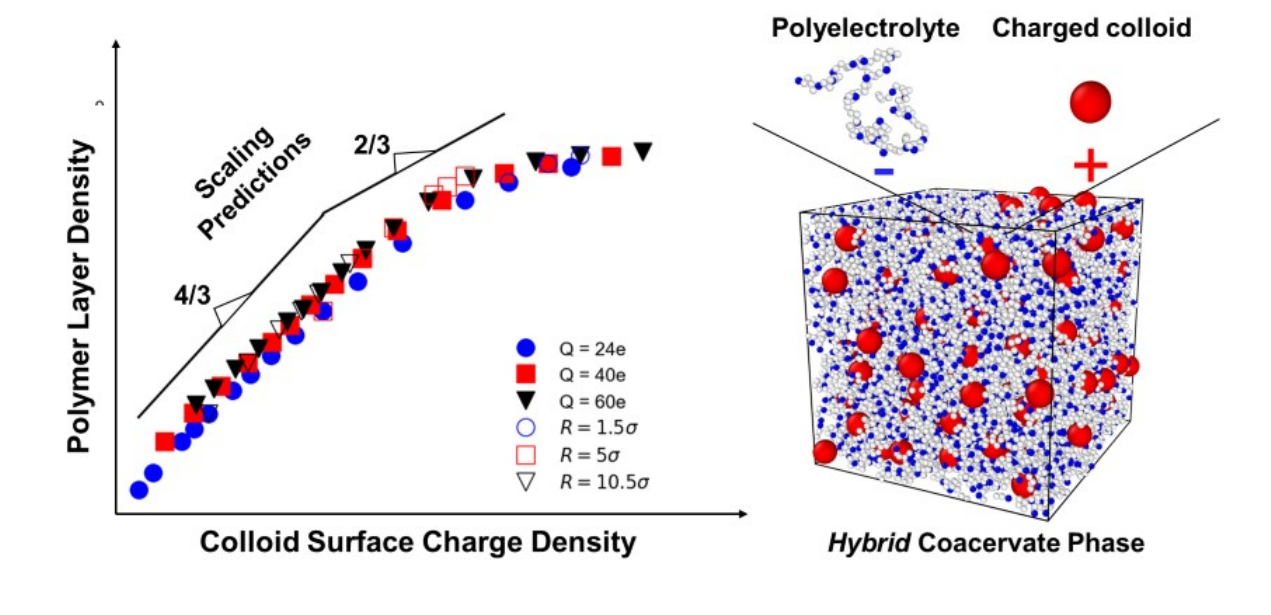}
\end{tocentry}

\begin{abstract}

Electrostatic interactions in polymeric systems are responsible for a wide range of liquid-liquid phase transitions that are of importance for biology and materials science. Such transitions are referred to as complex coacervation, and recent studies have sought to understand the underlying physics and chemistry.  Most theoretical and simulation efforts to date have focused on oppositely charged linear polyelectrolytes, which adopt nearly ideal-coil conformations in the condensed phase. However, when one of the coacervate components is a globular protein, a better model of complexation should replace one of the species with a spherical charged particle or colloid. In this work, we perform coarse-grained simulations of colloid-polyelectrolyte coacervation using a spherical model for the colloid. Simulation results indicate that the electroneutral cell of the resulting (hybrid) coacervates consists of a polyelectrolyte layer adsorbed on the colloid. Power laws for the structure and the density of the condensed phase, which are extracted from simulations, are found to be consistent with the adsorption-based scaling theory of coacervation. The coacervates remain amorphous (disordered) at a moderate colloid charge, $Q$, while an intra-coacervate colloidal crystal is formed above a certain threshold, at $Q > Q^{*}$. In the disordered coacervate, if $Q$ is sufficiently low, colloids diffuse as neutral non-sticky nanoparticles in the semidilute polymer solution. For higher $Q$, adsorption is strong and colloids become effectively sticky. Our findings are relevant for the coacervation of polyelectrolytes with proteins, spherical micelles of ionic surfactants, and solid organic or inorganic nanoparticles.

\end{abstract}

\maketitle

\newpage

\section{Introduction}

Over the past decade, complex coacervation between oppositely charged linear polyelectrolytes (PEs) has been studied extensively, by means of simulations, theory, and experiments. Substantial progress has been made in revealing the relationship between the coacervate properties and the molecular characteristics of the underlying PEs, such as their charge fraction, their monomer sequence, and their stiffness. In particular, theoretical and simulation work has rationalized experimental observations and provided valuable predictions regarding the equilibrium and dynamic properties of coacervates.~\cite{delaney2017theory,sing2020recent,rumyantsev2021polyelectrolyte,zhang2021interfacial} Much less is known, however, about coacervate systems where linear PEs are mixed with other charged colloidal species, such as charged solid nanoparticles, micelles of ionic surfactants, or globular proteins.

In this work, we present a systematic study of this type of coacervate. Specifically, we consider how substituting one PE component by charged colloids alters the properties of the resulting coacervate phases. We refer to these colloid-PE coacervate systems as \textit{hybrid} coacervates, to underscore the difference from conventional interPE coacervates, which are formed as the result of the electrostatically driven phase separation in solutions of oppositely charged PEs; this nomenclature also reflects the unique properties of colloid-PE complexes, which arise when the new, colloidal component is introduced. Interest in hybrid coacervates, particularly those formed by PEs and proteins, is partly motivated by the ability of PEs to stabilize the structure and functionality of proteins.~\cite{xu2017selective} There are promising implications for protein purification,~\cite{xu2011protein,xu2017selective} protein separation,~\cite{mctigue2019design,xu2017selective} antigen delivery,~\cite{hwang2020single,jeon2022enhancing} and food science, to name a few.~\cite{turgeon2009protein+,jun2011microencapsulation} Hybrid coacervates can also be viewed as model systems for intracellular compartmentalization, and for the formation of membrane-less organelles, particularly those comprising nucleic acids (DNA/RNA) and proteins that carry an opposite (i.e. positive) net charge.~\cite{devi2017encapsulation,blocher2020protein} Note that throughout this manuscript, the terms colloid and nanoparticle are used interchangeably.

Experimental studies of coacervates consisting of globular proteins and PEs have revealed that the formation of the coacervate phase can be influenced by many factors, including pH, which affects the net charge of the protein,~\cite{comert2016coacervation,comert2017liquid} the charge stoichiometry between proteins and PEs,~\cite{yan2013heteroprotein,kayitmazer2015complex} PE stiffness,~\cite{gao2019protein} and the distribution/patchiness of the (generally speaking, positive and negative) charges on the protein surface.~\cite{kim2020effect} Readers are referred to recent reviews~\citenum{comert2017liquid},~\citenum{kizilay2011complexation},~\citenum{horn2019macro}, and~\citenum{kapelner2021molecular} for a comprehensive overview of experimental work in this area. 

The similarities between hybrid and conventional coacervates are supported by available experimental data. For example, increasing the charge density that controls the strength of electrostatic interactions has been shown to promote coacervation for both interPE and colloid-PE coacervates. For conventional coacervates of linear PEs, an increase in the fraction of ionic monomers at fixed chain length, i.e. an increase in the PE total charge, leads to denser coacervate phases and to higher salt resistance.~\cite{neitzel2021expanding} Similarly, for hybrid coacervates of green fluorescent protein (GFP) and synthetic or biological polyanions, Cummings and Obermeyer have found that an increase of the net charge of GFP from $Q = +2e$ to $Q = + 26e$ facilitates coacervation over a wider range of pH and salt concentration.~\cite{cummings2018phase} 
Initial observations were collected \textit{in vitro}, but similar trends
also hold \textit{in vivo} for \textit{E. Coli}, where the supercharged GFP with a sufficiently high charge forms intracellular condensates via complexation with RNA.~\cite{yeong2020formation}

The 1D sequence of neutral and ionic monomers in linear PEs and the 2D distribution of charges on a protein's surface also influence the phase behaviors of conventional/linear and hybrid coacervates. For coacervates of linear PEs, the monomer sequence has been shown to govern coacervation in a way where a higher blockiness of charged monomers favors the formation of denser and more salt-resistant coacervate phases.~\cite{rumyantsev2019controlling, yu2021complex, Sing-2017, lytle2019designing} Charge patchiness on a protein's surface has an analogous effect on the formation of hybrid coacervates. Comparison across a series of GFP mutants demonstrated that, upon hybrid coacervation with various linear synthetic polyanions, proteins that exhibit a higher anisotropy of surface charge form condensed phases, which remain stable up to higher salt concentrations.~\cite{kim2020effect}

Computational and theoretical studies of hybrid coacervates have been limited. A recent report ~\citenum{xiao2017application} provides an overview of how Monte Carlo simulations combined with a single chain in mean field methodology can be used to investigate the structure of PE-charged particle mixtures. Within this approach, one can outline the conditions necessary for the formation of the macroscopic condensed phase or the finite-size aggregates of nanoparticles with PEs. A similar approach has been applied by Ganesan and co-workers to demonstrate the role of surface charge patchiness on the coacervation between proteins and PEs.~\cite{samanta2018influence,samanta2020influence,samanta2020direct} Madinya and Sing considered the phase behavior of hybrid coacervates between worm-like ionic micelles and PEs using a hybrid Monte Carlo and self-consistent field theory model.~\cite{madinya2022hybrid} Taken together, these works have provided helpful insights into the equilibrium complexation between PEs and colloids/surfactants.  

To arrive at more universal, analytical considerations on hybrid coacervates and in an effort to extend theoretical predictions into their dynamic and rheological behavior, we have recently developed a scaling theory of coacervation between linear PEs and colloids.~\cite{artem2022hybrid} In that approach, colloidal nanoparticles are treated as charged and impenetrable spheres, and hybrid coacervation with PEs is treated as the adsorption of flexible PE chains at the charged spheres followed by bridge-driven association of the resulting neutral complexes. This approach enabled prediction of the dependence of hybrid coacervates' structural properties, such as the average polymer density and the thickness of the PE layer between adjacent colloids, and their dynamic properties, such as the coacervate viscosity and colloid diffusion coefficient, on the size and net charge of the colloid. Building on this theoretical analysis, in this work we seek to test our scaling predictions and assumptions, and provide molecular-level insights to will further improve and refine existing theoretical arguments. Moreover, simulations allow us to go beyond the limits of the scaling theory applicability, and explore regions of parameter space that have not been described theoretically, and where new and unexpected behaviors may occur.

We rely on coarse-grained molecular dynamics simulations based on the Kremer-Grest model~\cite{kremer1990dynamics}, augmented by Coulomb interactions. That model has been applied  successfully to model both the structural~\cite{yu2021complex,andreev2018complex,bobbili2021simple} and rheological properties~\cite{yu2020crossover,liang2022coarse} of conventional coacervates. To examine hybrid systems, we replace one type of PE chain with charged spherical particles. Consistent with our theoretical representation~\cite {artem2022hybrid}, we model a colloid particle as a spherical interaction site with the charge either assigned to the center of the sphere or uniformly smeared over the sphere's surface. It is important to note that, for equal size and net charge of the colloid, these two alternatives appear to be equivalent and lead to identical results. We limit our simulations to salt-free hybrid coacervates and focus on the dependencies of the coacervate structural and dynamical properties on the nanoparticle radius $R$, charge $Q$, and PE chain length $N$. Simulation results are systematically compared to theoretical scaling (power) laws derived in ref.~\citenum{artem2022hybrid}.

This manuscript is organized as follows. The details of our coarse-grained model are described in Section~\ref{section:method}. Sections~\ref{section:structure}-\ref{section:dynamics} present simulation results. In Section~\ref{section:structure}, we start by testing the scaling dependence of the structure of coacervates, namely the average polymer density and the thickness of the PE layer surrounding each colloid, on the characteristics of charged nanoparticles, $Q$ and $R$. Section~\ref{section:modulus} discusses how the bulk modulus of the hybrid coacervate is affected by particle size and charge. To provide insights into the dynamics of colloids within the hybrid coacervate, their diffusion is considered in Section~\ref{section:dynamics}. Particular attention is paid to the effect of chain length, which triggers a Rouse-to-reptation crossover in the PE chain dynamics~\cite{yu2020crossover}, and governs the polymer-mediated mobility of the nanoparticle. The findings of this work are summarized in Section~\ref{section:conclusion}.

\section{Simulation Methods}
\label{section:method}

Each PE chain is represented by a set of spherical interaction sites (beads) connected by springs.~\cite{kremer1990dynamics} Each chain has the same charge fraction, $f = 0.2$, which is the ratio between the number of charged beads and the chain length, $N$. Since previous studies~\cite{Sing-2017,lytle2019designing,rumyantsev2019controlling} have demonstrated that the charge sequence greatly influences the phase behavior of coacervates, the charged beads are equidistantly distributed along each PE chain. The colloids are modeled as spherical particles with radius $R$,~\cite{Liu2011, Liu2008} net charge $Q$, and unit mass. Within the first representation of the colloid, for each particle the charge $Q$ is placed at the center of a sphere. In the second representation, we implement a uniform smearing of the charge throughout the sphere's surface, where $n$ monomers, each carrying charge $Q/n$ and mass $0.5/n$, are evenly distributed on a sphere of mass $0.5$. At high $n$, the external electric field around the spheres created through the first or the second charge configurations is almost identical. This leads to similar results for the hybrid coacervate properties, as discussed in Section 1 of the Supporting Information. For this reason, and for simplicity, we employ the first representation in most of our simulations. In our simulations, the polymers are under $\Theta$ solvent conditions, and an implicit solvent is adopted for computational efficiency.

The connectivity of copolymer chains is described by a 
finitely extensible nonlinear elastic (FENE) potential between bonded beads:
\begin{equation}
U_{FENE}=
-0.5KR_{0}^{2} \ln \left[ 1 - \left( \dfrac{r}{R_0} \right)^2 \right]
\end{equation} 
with $K=30k_{B}T/\sigma^{2}$ and $R_{0} = 1.5\sigma$.~\cite{kremer1990dynamics}
All beads interact through a shifted and truncated Lennard-Jones (LJ) potential:
\begin{equation}
\label{LJ-c5}
U_{LJ}=\left\{ \begin{array}{lcl}
4 \varepsilon \left[ \left( \dfrac{\sigma_{i}}{r} \right)^{12} - \left( \dfrac{\sigma_{i}}{r} \right)^{6}-\left( \dfrac{\sigma_{i}}{r_c} \right)^{12} + \left( \dfrac{\sigma_{i}}{r_c} \right)^{6}\right] & \mbox{for}
& r\leq r_c  \\
0 & \mbox{for} & r > r_c
\end{array}\right.
\end{equation}
where $\varepsilon$ describes the strength of interaction, $r_c$ is the cutoff radius, and $\sigma_i$ is the bead diameter; here we use $i = m$ for the monomer bead, $i = p$ for the charged colloid particle, and $i = mix$ for the monomer-colloid pair. Namely, $\sigma_{m} = \sigma$, $\sigma_{p} = 2R$ with $R$ equal to the radius of the particle, and $\sigma_{mix} = \sigma_{p} / 2 + \sigma$ is defined using a mixing rule between the monomer bead and the colloid particle. For all bonded beads, $\varepsilon = k_{B}T$, $\sigma_{i} = \sigma_{m}$, and $r_{c} = 2^{1/6}\sigma_{m}$ to balance the attraction provided by the FENE potential and maintain a nonzero length of the bond. For all non-bonded monomer beads, $\varepsilon = 0.314 k_{B}T$, $\sigma_{i} = \sigma_{m}$, and $r_{c} = 2.5\sigma_{m}$ to represent $\Theta$ solvent conditions.~\cite{graessley1999excluded, neitzel2021polyelectrolyte} The impenetrability of the colloid particles is enforced by $\varepsilon = k_{B}T$, $\sigma_{i} = \sigma_{p}$, and $r_{c} = 2^{1/6}\sigma_{p}$ for the LJ potentials between them. The parameters corresponding to LJ interactions between monomer beads and colloids are $\varepsilon = k_{B}T$, $\sigma_{i} = \sigma_{mix}$, and $r_{c} = \sigma_{mix}$.
The electrostatic interactions in the system are given by
\begin{equation}
\frac{U_{coul}}{k_{B}T}=\frac{z_{i}z_{j}l_{B}}{r}
\end{equation}
where $z_{i}$ is the charge valence for species $i$: $z_{m} = -e$ for charged monomers and $z_{p} = +Q/e$ for the colloid. The Bjerrum length $l_{B} = e^2 / \epsilon k_B T$ is set to $l_{B} = \sigma$ in this work. Coulomb interactions are computed by the particle-particle particle-mesh (PPPM) method, with the error for the long-range force set to be within $10^{-4}$.

To simulate a salt-free coacervate phase in the equilibrium state, the simulation box is maintained in an NPT ensemble with external pressure $P = 0$. This corresponds to approximately zero osmotic pressure of the highly diluted supernatant coexisting with the coacervate.~\cite{rubinstein2018structure,yu2020crossover} This NPT ensemble is achieved by coupling a Berendsen barostat and a Langevin thermostat with damping parameter $\Gamma = 1.0m/\tau_{LJ}$, where $\tau_{LJ}$ is the reduced LJ time unit and $m = 1$ is the reduced particle mass. Bead velocities and positions are updated by a velocity-Verlet algorithm. The time step is set to $0.01\tau_{LJ}$. The equilibrium of the system is ensured by monitoring the convergence of the coacervate density and the relaxation of end-to-end auto-correlation vectors of the PE chains.~\cite{yu2020crossover} All the properties of the coacervate phase are measured by block averaging after the systems reach equilibrium. Simulation snapshots are generated using the OVITO software.~\cite{ovito}

\section{Structural Properties of Colloid-Polyelectrolyte Coacervates}
\label{section:structure}

In this Section, we focus on the effects of particle radius $R$ and net charge $Q$ on the properties of the hybrid coacervate phase. In Sections \ref{section:structure} and \ref{section:modulus} that deal with structural properties, we limit our simulations to the charge-matched case when the charge on the PE chain is equal to the colloid charge, $Q/e = fN$. Theory suggests that, for a fixed $Q$ value, the coacervate structure should remain unchanged for any PEs of higher length, $fN > Q/e$.~\cite{artem2022hybrid} 

\begin{figure}[ht]
\centering
\includegraphics[width=0.8\linewidth]{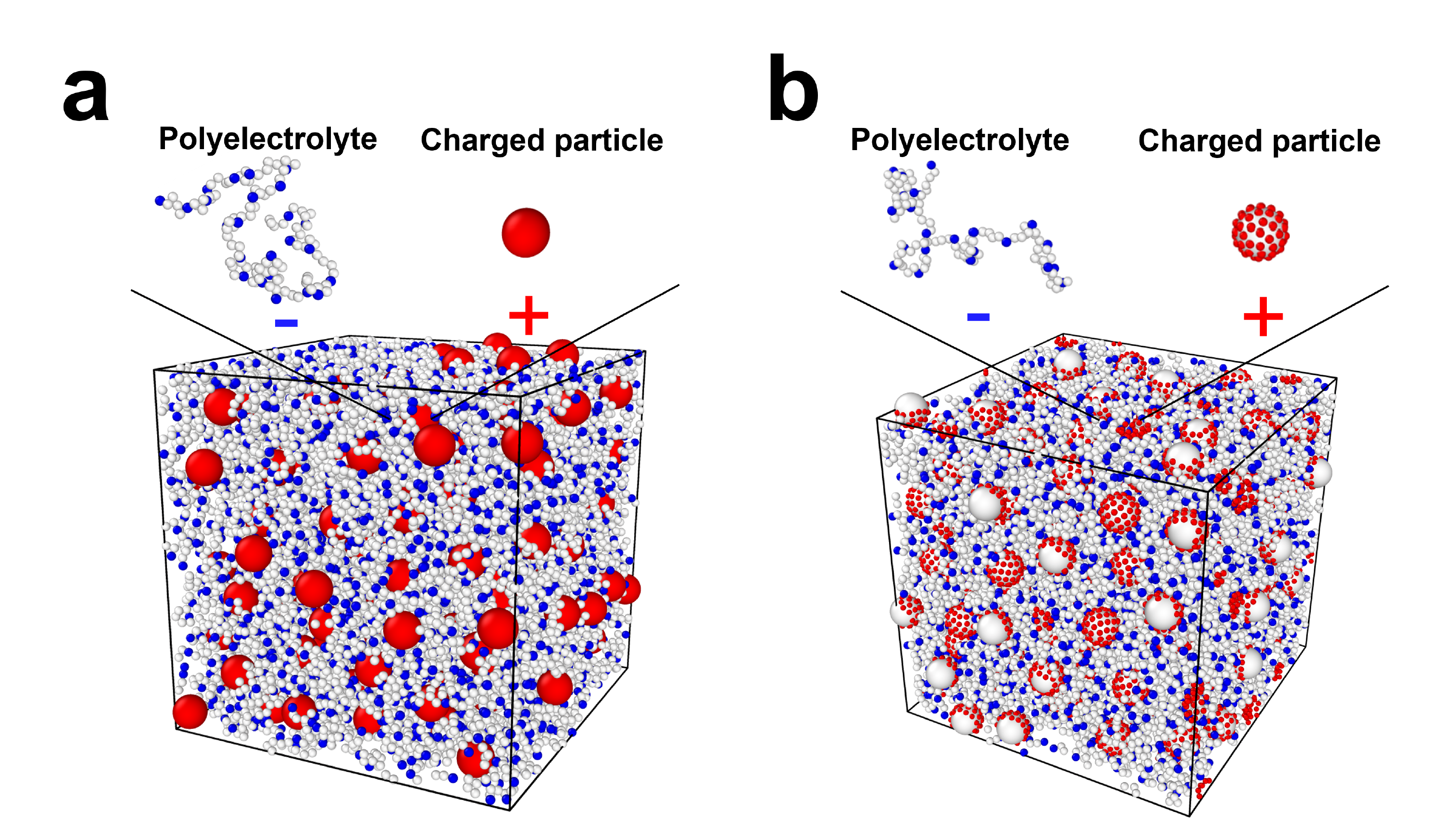}
\caption{ Representative snapshots of the hybrid coacervate phase in simulations with a) the charge $Q$ assigned to the center of each particle; 
b) the charge uniformly distributed over the nanoparticle surface ($64$ charged sites are shown in red, each carrying the fractional charge of $3/8e$). A detailed comparison can be found in the Supporting Information. The simulation parameters are given by $f = 0.2, R = 2\sigma, Q = 24e$, and $l_{B} = \sigma$. Blue, red, and white beads represent negatively charged, positively charged, and neutral beads (sites).}
\label{fig:snapshots-c5}
\end{figure}

Representative snapshots of the salt- and counterion-free hybrid coacervate phase are shown in Figure~\ref{fig:snapshots-c5}. For both implementations of the colloid nanoparticles --- with a single $Q$-charge at their centers, as shown in Figure~\ref{fig:snapshots-c5}a, and for the charge uniformly distributed over the particle surface, as shown in Figure~\ref{fig:snapshots-c5}b --- one can see that the hybrid coacervate is a homogeneous mixture of charged particles and linear PE chains. PE chains adsorb onto particles and form bridges that connect neighboring colloids. Since the two approaches provide quantitatively identical results (see Section 1 in the Supporting Information), as already noted,
in the remainder of this work the colloidal charge is modeled by assigning a single $Q$-charge to the particle center. 

The scaling theory of the hybrid coacervate was introduced in ref.~\citenum{artem2022hybrid}, and we refer to this work for detailed derivations. In what follows, we briefly summarize the key theoretical predictions and conclusions regarding the structure of the colloid-PE coacervates, which hold for $Q/e \neq fN$. The key assumption in the theory is the absence of ion pairing between the charges in the PE and on the colloid, given that the radius of the nanoparticles is sufficiently larger than the monomer size. 

The theory considers an elementary electro-neutral cell of the hybrid coacervate, which consists of the colloid nanoparticle and the PE chains, as the PE adsorbs on the oppositely charged colloid. Therefore, the structure of the hybrid coacervate phase is given in terms of the structure of the absorbed PE layer for each electroneutral cell. The central theoretical predictions are the scaling dependencies of the polymer volume fraction within the absorbed layer, $\phi$, and its thickness, $H$, on nanoparticle radius and charge, $R$ and $Q$. Depending on the strength of the absorption, which is classified as strong or weak, and the geometry of the adsorbed PE layer, which is either essentially spherical (for $H \gg R$) or quasi-planar for ($R \gg H$), theory distinguishes several different adsorption regimes.  We note that the strength of adsorption is defined by the type of repulsive interactions that balance Coulomb attractions between colloids and PEs.~\cite{artem2022hybrid} If they are short-range three-body repulsions, adsorption is considered strong. If repulsions originate from the PE conformational entropy, adsorption is classified as weak. Ref.~\citenum{artem2022hybrid} predicts that there are three different scaling regimes for the structure of the hybrid coacervate phase, which correspond to i) the strong spherical absorption, Regime I; ii) the strong quasi-planar absorption, Regime II; iii) the weak quasi-planar absorption, Regime III. The following scaling laws are predicted to describe the hybrid coacervate structure in Regime I:
\begin{equation}
\phi_{I} \simeq u^{3/5}f^{4/5}Q^{2/5}
\label{eq:region1-phi}
\end{equation}
\begin{equation}
H_{I} \simeq u^{-1/5}f^{-3/5}Q^{1/5}
\label{eq:region1-H}
\end{equation}
Here $u = l_B/  a$ is the theoretical dimensionless parameter equal to the ratio between the Bjerrum length $l_B$ and the statistical segment size $a$. For Regimes II and III, the respective laws can be written as
\begin{equation}
\label{eq:region2-phi}
\phi_{II} \simeq u^{1/3} \left( \frac{Q}{R^2} \right)^{2/3} \simeq u^{1/3}Q^{2/3}R^{-4/3}
\end{equation}
\begin{equation}
\label{eq:region2-H}
H_{II}  \simeq u^{-1/3} \left( \frac{Q}{R^2} \right)^{1/3} \simeq u^{-1/3}f^{-1}Q^{1/3}R^{-2/3}
\end{equation}
\begin{equation}
\label{eq:region3-phi}
\phi_{III}  \simeq u^{1/3} \left( \frac{Q}{R^2} \right)^{4/3} \simeq u^{1/3}f^{-2/3}Q^{4/3}R^{-8/3}
\end{equation}
\begin{equation}
\label{eq:region3-H}
H_{III}  \simeq u^{-1/3} \left( \frac{Q}{R^2} \right)^{-1/3} \simeq u^{-1/3}f^{-1/3}Q^{-1/3}R^{2/3}
\end{equation}
Eqs.~\ref{eq:region2-phi}-\ref{eq:region3-H} show that the properties of the hybrid coacervate in the regimes of quasi-planar adsorption, II and III, are controlled by the surface charge density of the colloid, equal to $Q/R^2$. This is in contrast to Regime I, where $\phi_{I}$ and $H_{I}$ are functions of the colloid charge but not its radius.

As the particle radius $R$ increases or its net charge $Q$ decreases, the hybrid coacervate undergoes a continuous crossover from Regime I to Regime II and then from Regime II to Regime III. The reverse order of transitions can be triggered by the decrease of $R$ or the increase of $Q$.~\cite{artem2022hybrid}

\subsection{Density of the Polymer Layer}
\label{subsection:phi}

We start by analyzing the dependence of the density of the absorbed PE layer, $\phi$, on the colloid radius and charge, $R$ and $Q$. It should be noted that $\phi$ quantifies the average density of the PE layer, rather than the average polymer density of the entire hybrid coacervate. Therefore, in our simulations, $\phi$ is calculated as the average monomer number density within the volume occupied by the PEs: 
\begin{equation}
\phi = \frac{N \mathcal{N}_{c} \sigma^{3} }{l^{3} - 4/3\pi R^{3} \mathcal{N}_{p} }  
\label{phi-definition}
\end{equation}
Here $ \mathcal{N}_{c}$ is the total number of PE chains in the system, $N$ is the PE chain length, $l$ is the length of the cubic box, and $\mathcal{N}_{p}$ is the total number of colloids. In eq.~\ref{phi-definition}, the denominator is equal to the volume of the hybrid coacervate occupied by PEs, which is the difference between the total volume of the simulation box, $l^3$, and the volume of all colloidal particles. The number density $\phi$ is calculated according to eq.~\ref{phi-definition} and is proportional to the theoretical polymer volume fraction given by eqs.~\ref{eq:region1-phi}, \ref{eq:region2-phi}, and \ref{eq:region3-phi}; these quantities exactly coincide if the monomer volume equals $\sigma^3$. This justifies the comparison of the simulation results to the theoretical scaling laws.

\begin{figure}[ht]
\centering
\includegraphics[width=0.8\linewidth]{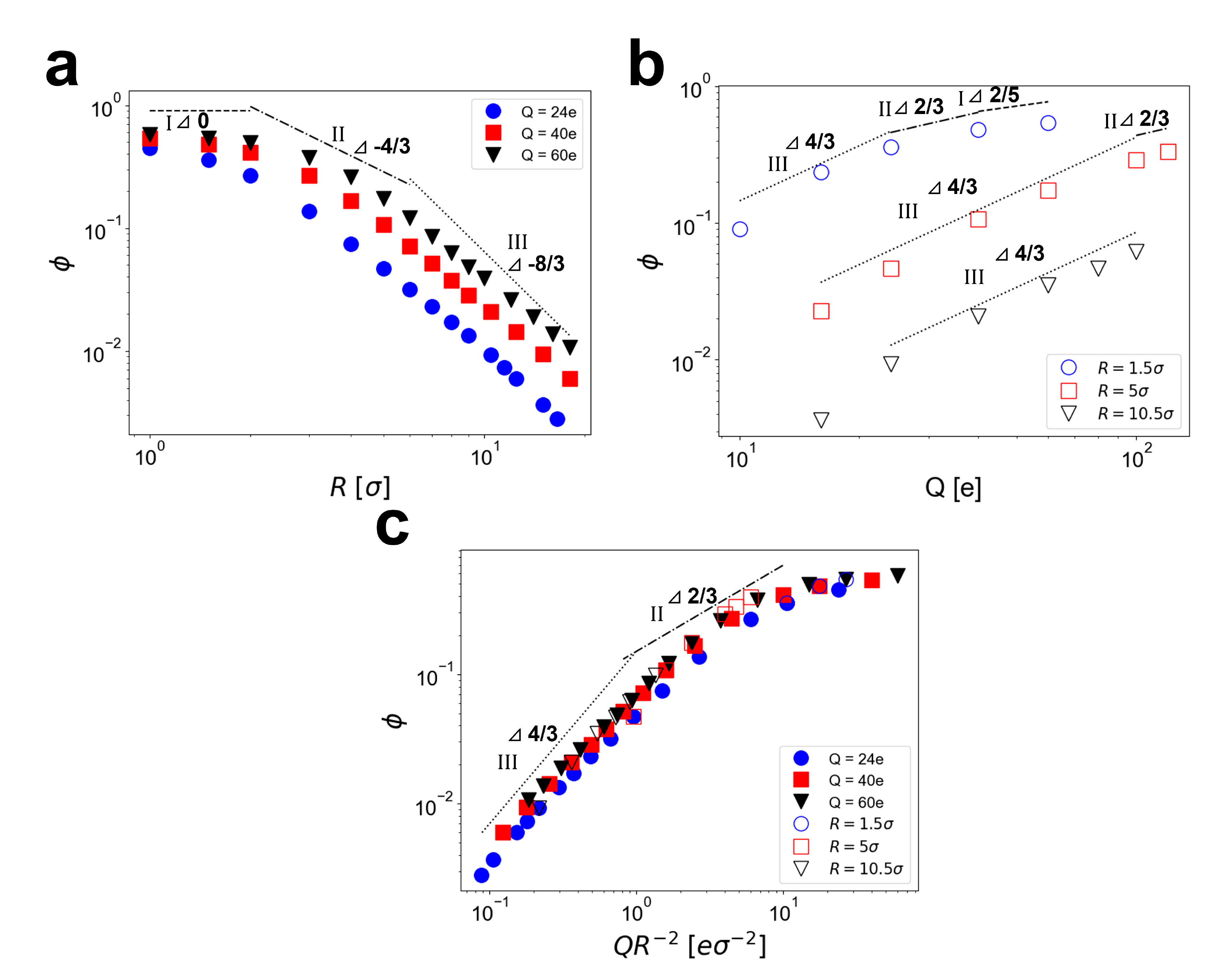}
\caption{Density of the adsorbed PE layer $\phi$ as a function of 
a) the particle radius $R$ for $Q = 24e$, $40e$, and $60e$;
b) the particle charge $Q$ for $R = 1.5 \sigma$, $5 \sigma$, and $10.5 \sigma$.
c) the surface charge density of the colloid, $Q / R^2$, for $Q = 24e, 40e$, and $60e$.
The data points represent simulation results. The errors correspond to the standard deviation and do not exceed the size of the symbols. All results are shown on a log-log scale, and the straight lines represent the theoretical predictions for the relevant scaling regimes. The numbers on the right of the triangular symbol show the values of the scaling exponents (slopes). The simulation parameters are set to $f = 0.2$, and $l_{B} = \sigma$, and colloid-PE pairs are charge-matched, $Q/e=fN$. }
\label{fig:phi}
\end{figure}

The PE layer density $\phi$ as a function of the particle radius $R$ is shown in Figure~\ref{fig:phi}a. Three sets of data points correspond to different net charges of the colloid, $Q$. All results are plotted on a log-log scale so that the theoretical laws given by eqs.~\ref{eq:region1-phi}-\ref{eq:region3-H} are shown with straight lines. The $\phi(R)$ dependencies obtained in simulations closely follow the scaling predictions, and can be indeed classified into three regions for the predicted scaling regimes, I-III. 

For small colloid radii, $R \leq 2 \sigma$, a plateau can be observed for $Q = 60e$, in agreement with eq.~\ref{eq:region1-phi} suggesting no dependence of the PE layer density on the colloid radius in Regime I. For a lower net charge of the colloid, $Q = 40e$ and $Q = 24e$, this plateau vanishes, which is consistent with the shift of the I/II crossover to lower $R$ values at decreasing $Q$~\cite{artem2022hybrid} 
\begin{equation}
R_{I/II} \simeq u^{-1/3} f^{-3/5} Q^{1/5}
\label{crossover-I/II}
\end{equation}
This result for the crossover can be derived by comparing eqs.~\ref{eq:region1-phi} and \ref{eq:region2-phi}. For larger $R$, an intermediate regime with a slope close to $-4/3$ can be seen for all curves, which is consistent with scaling Regime II. A further increase of the colloid radius $R$ leads to a much faster decrease in the density $\phi$, with the slope asymptotically close to $-8/3$; the latter slope is theoretically anticipated in Regime III. Thus, the observed continuous increase of (the absolute value of) the scaling exponent for increasing $R$, which accompanies the crossovers I/II and II/III, is consistent with theoretical conclusions of ref.~\citenum{artem2022hybrid}.

It is of interest to compare the hybrid coacervates' properties across the different values of the colloid charge, $Q$. First, Figure~\ref{fig:phi}a shows that, as $Q$ increases, the entire density curve shifts upwards. This indicates that the increasing colloid charge facilitates the formation of the denser absorbed layers, given that the colloid radius remains unchanged. In other words, increasing the strength of Coulomb interactions between the colloid and the PE promotes hybrid coacervation. Second, as $Q$ increases, the transition between the different scaling regimes becomes more evident and Regime II becomes noticeably wider. The positions of the I/II and II/III crossovers shift to a larger $R$. For $Q = 24e$, the dependence of $\phi$ on $R$ almost immediately tends to $-8/3$, without a clear region of slope $0$ or slope $-4/3$. In contrast, for $Q = 60e$, a plateau develops up until $R = 2\sigma$, and the dependence of $\phi$ on $R$ reaches a slope of $-8/3$ at a much larger $R$ compared with the case of $Q = 24e$. In addition, the window of with an intermediate slope of $-4/3$ is much clearer for higher $Q$ values. The detected shift of the crossovers agrees with theoretical predictions given by
\begin{equation}
R_{II/III} \simeq f^{-1/2} Q^{1/2}
\label{crossover-II/III}
\end{equation}
and eq.~\ref{crossover-I/II}. The scaling theory suggests that the boundary $R$ values increase with $Q$ for both crossovers.~\cite{artem2022hybrid} Moreover, the $R$-width of Regime II can be estimated as
\begin{equation}
\Delta R_{II} \simeq \frac{R_{II/III}} {R_{I/II}} \simeq u^{1/3} f^{1/10} Q^{3/10}
\end{equation}
and increases with $Q$. This explains why, in our simulation results, the intermediate Regime II is much better delineated for highly charged colloids. 

To examine the effect of colloid charge on the density of the PE layer in a more systematic manner, we present the respective dependence in Figure~\ref{fig:phi}b. Again, the slopes of the straight lines reflect the theoretical predictions of eqs.~\ref{eq:region1-phi}, \ref{eq:region2-phi}, and \ref{eq:region3-phi} for the $\phi(Q)$ dependencies in Regimes I, II, and III. The simulation results show a change of the scaling exponent at increasing $Q$ and approximately follow the theoretical expectations. However, the difference between Regime II, where the slope is equal to $2/3$, and Region I with a slope of $2/5$ is difficult to distinguish. The fact that the apparent slope for the simulation results continuously decreases at increasing $Q$ is consistent with the scaling, which predicts the crossover from the Regime III of weak adsorption to the strong adsorption regime, Regime II, and then to Regime I. The scaling exponent is predicted to change from $4/3$ to $2/3$ and then to $2/5$. The monotonic increase of the density $\phi$ with colloid charge $Q$ is also qualitatively consistent with the experimental observations.~\cite{cummings2018phase}

Similar to what was discussed earlier for $\phi (Q)$, the positions of the III/II and II/I crossovers in the $\phi (Q)$ dependence shift to higher $Q$ values as $R$ increases. As a result, only for $R = 1.5 \sigma$, all three regimes can be presumably distinguished in the simulation data, but each regime spans a very narrow $Q$ range. For larger colloids, such as that with $R = 10.5\sigma$, only Region III can be convincingly identified in the range of the $Q$ values considered in our simulations. By inverting eqs.~\ref{crossover-I/II} and \ref{crossover-II/III}, one can see that the shift of the crossover position across different $R$ values agrees well with the scaling picture: Theoretically, these crossovers can be written as $Q_{III/II} \simeq f R^2$ and 
$Q_{II/I} \simeq u f^3 R^5$. This strong increase in the crossover $Q$ values helps explain why Regime II is barely distinguishable and Regime I is absent for bulky colloids. Another limitation, which precludes us from sampling the region of high $Q$ values and detecting Regimes II and I for bulky colloids, is the formation of a colloidal crystal, whereas scaling predictions were obtained for liquid/amorphous hybrid coacervates only. This intriguing phenomenon is briefly discussed in Subsection~\ref{crystal}. The additional challenge pertains to the very slow equilibration of even an amorphous hybrid coacervate at high colloid charges.

According to eqs.~\ref{eq:region2-phi} and \ref{eq:region3-phi}, for the regimes of quasi-planar adsorption, II and III, the dependence of the polymer layer density $\phi$ on particle charge and radius can be reduced to that of the single quantity, $Q / R^2$, which corresponds to the surface charge density of the colloids. Namely, scaling laws read 
$\phi_{II} \simeq (Q / R^2 )^{2/3}$ for Region II and $\phi_{III} \simeq (Q / R^2 )^{4/3}$ for Region III. Therefore, if the simulation data for the polymer layer density $\phi$
are represented in the coordinates of the surface charge density $Q /R^{2}$, the results should collapse onto a master curve over the entire range of Regimes II and III. Figure~\ref{fig:phi}c shows that the simulation results support this scaling universality idea, and the respective slopes for Regimes II and III are in good agreement with the theoretical values of $2/3$ to $4/3$. 


In addition to the power laws, simulations also allow us to test the radial density profile of the PE layer that coats the colloids in the condensed phase. Scaling theory predicts that, in Regime I of the essentially spherical adsorption, the internal structure of the adsorbed PE layer is inhomogeneous (see Appendix A of ref.~\citenum{artem2022hybrid}). Because Coulomb attractions between the colloid and the distal part of the PE layer are partially screened by the internal part of this layer, the density of the polymer layer is the highest near the surface of the colloid and decreases with increasing distance from it. To corroborate this theoretical result, the average monomer number density within the thin spherical shell is calculated as a function of the distance between the particle center and the shell. The resulting radial density profile of the PE layer, which was obtained for $Q = 40e$ and $R = 1.5\sigma$ and therefore corresponds to the scaling Regime I, is shown in Figure S2 of the Supporting Information. It shows that the density near the colloid surface is indeed approximately 40\% higher than at the periphery of the PE layer, in qualitative agreement with the scaling picture of the hybrid coacervate.~\cite{artem2022hybrid}

\subsection{Polymer Layer Thickness $H$}

In our simulations, the thickness of the absorbed PE layer coating each colloid within the hybrid coacervate is determined as follows:
\begin{equation}
H = \frac{ \Delta d_{p} }{2} - R
\end{equation}
Here $\Delta d_{p}$ is the average distance between two neighboring nanoparticles, and $R$ is the particle radius; $\Delta d_{p}$ can be identified as the position (radial coordinate) of the first peak in the particle-particle radial distribution function (RDF). A representative RDF is shown in Figure S3a of the Supporting Information. 

\begin{figure}[ht]
\centering
\includegraphics[width=0.75\linewidth]{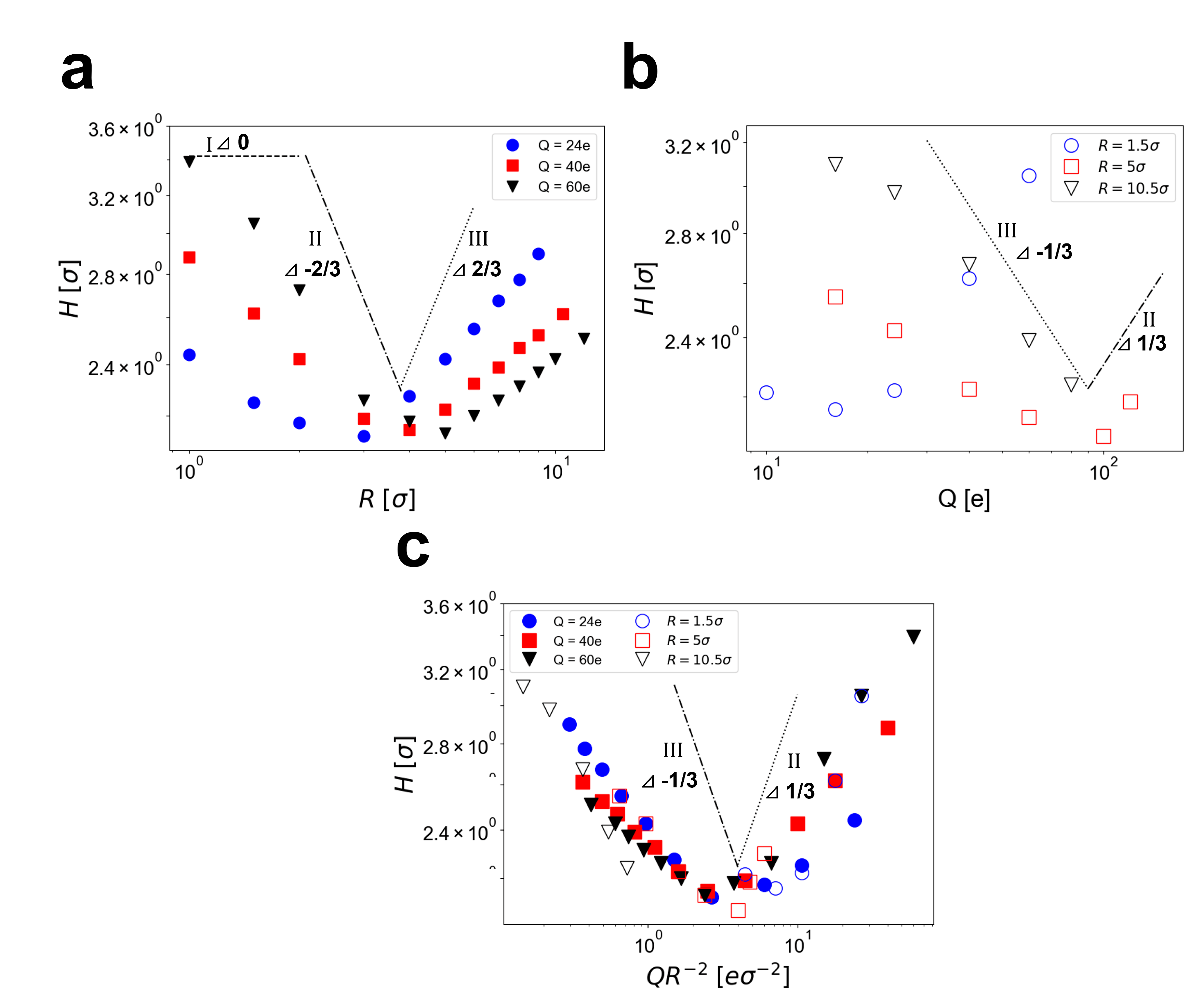}
\caption{Polymer layer thickness $H$ as a function of
a) the particle radius $R$ for the net colloid charge $Q$ = $24e$, $40e$, and $60e$; 
b) the particle charges $Q$ for colloid radius $R = 1.5 \sigma$, $5\sigma$, and $10.5\sigma$;
c) the surface charge density of the colloid, $Q / R^2$, for $Q = 24e$, $40e$, and $60e$.
The data points represent simulation results obtained by averaging over several independent runs. The errors are estimated by the standard deviation and are within the size of the symbols. All dependencies are shown in a log-log scale, and the straight lines represent the theoretical scaling laws with the numbers indicating their slopes (exponents). Hybrid coacervates are charge-matched, $Q = f N$, and the simulation parameters are set to $f = 0.2$ and $l_{B} = \sigma$.}
\label{fig:thickness}
\end{figure}

Figure~\ref{fig:thickness}a demonstrates how the resulting thickness of the layer, $H$, changes with the colloid radius $R$ for the three values of the colloid charge, $Q = 24e, 40e$, and $60e$. Perhaps the most interesting finding is that the $H (R)$ dependence is non-monotonic, with the thickness decreasing at a small $R$, and then increasing when colloids become sufficiently large. Remarkably, this non-monotonic was predicted within the scaling considerations of ref.~\citenum{andreev2018complex}, and the respective power laws are shown in Figure~\ref{fig:thickness}a with straight lines. The physical reason for the non-monotonic behavior stems from the competition between two factors. On the one hand, increasing $R$ would make the thickness of the layer smaller if the total layer density (and hence volume) remained unchanged. On the other hand, increasing the colloid size weakens the Coulomb attractions with the PE, and the average layer density goes down. The first, geometric factor prevails in Regime II of quasi-planar strong adsorption. Here the average density of the layer decreases, $\phi_{II} \sim R^{-4/3}$ according to eq.~\ref{eq:region2-phi}, but the geometry of the quasi-planar adsorption nevertheless leads to the decreasing $H$, $H_{II} \sim \left( \phi_{II} R^2 \right)^{-1} \sim R^{-2/3}$. As the colloid radius grows, the system enters Regime III of weak quasi-planar adsorption, where the second tendency becomes even stronger and takes over the first one: Eq.~\ref{eq:region3-phi} suggests that $\phi_{III} \sim R^{-8/3}$ so that the layer thickness increases, $H_{III} \sim \left( \phi_{III} R^{-2} \right)^{-1} \sim R^{2/3} $.

The simulations do not reproduce quantitatively the predicted exponents for the $H(R)$ dependence, which should be primarily attributed to the presence of the II/III crossover and a low magnitude of thickness change. The latter does not exceed $50\%$ whereas density changes approximately 10 times in the same range of $R$, thereby enabling easier testing of the respective scaling slopes. As in Figure~\ref{fig:phi}a, the theoretically expected plateau of Regime I is not seen in Figure~\ref{fig:thickness}a at low $R$ values. Much higher charge values would be needed to test this regime, which are not accessible in simulations due to colloidal crystallization.  

The simulation results shown in Figure~\ref{fig:thickness}a also demonstrate the shift of the II/III crossover positions, i.e., the position of the minimum in the $H (R)$ dependence, at increasing $Q$. This shift is quantitatively consistent with eq.~\ref{crossover-II/III} implying that $R_{II/III} \sim Q_{II/III}^{1/2}$. As the charge increases from $Q = 24e$ to $Q = 60e$, the minimum shifts from $R = 3 \sigma$ to $R= 5 \sigma$, i.e. its value increases $5/3 = 1.67$ times. The theoretical ratio $(60 / 24)^{1/2} = 1.58$ is very close to what is observed in simulations.

The dependence of $H$ on the colloid charge $Q$ is also non-monotonic, as seen in Figure~\ref{fig:thickness}b. The minimum in this dependence can be attributed to the competition of the same two factors and is also consistent with the scaling prediction of $H_{III} \sim Q^{-1/3}$ in Regime III but $H_{II} \simeq Q^{1/3}$ in Regime II. As $R$ increases, the crossover II/III shifts to higher $Q$. This is in line with the theoretical prediction of eq.~\ref{crossover-II/III}, which can be written as $Q_{II/III} \sim R^2$. As R increases from $1.5\sigma$ to $5\sigma$, the corresponding minimum changes from $Q = 10e$ to $Q = 100e$, i.e. increases 10 times. This agrees well with the scaling estimate $(5/1.5)^{2} \approx 11$.

Since the shift of the II/III crossover is well described by the scaling theory, all simulation data for Regimes II and III should collapse onto a universal master curve when plotted as a function of the reduced coordinate of $Q/R^2$, which is the surface charge density of the colloid. The physical reason for that is the quasi-planar geometry of Regimes II and III. Figure~\ref{fig:thickness}c shows that the expected universality is apparent, and the master curve exhibits the universal position of the minimum. It should be noted that the non-monotonic dependence of the height of the PE layer on the surface charge density was also predicted by Dobrynin, Deshkovskii, and Rubinstein in the context of PE adsorption at planar oppositely charged surfaces.~\cite{dobrynin2000,dobrynin2001} This problem is analogous to the Regimes II and III of quasi-planar PE adsorption in hybrid coacervates when the layer thickness is much smaller than the colloid radius, $H \ll R$.~\cite{artem2022hybrid}

\subsection{Formation of Colloidal (Wigner) Crystal}
\label{crystal}

\begin{figure}[ht]
\centering
\includegraphics[width=0.75\linewidth]{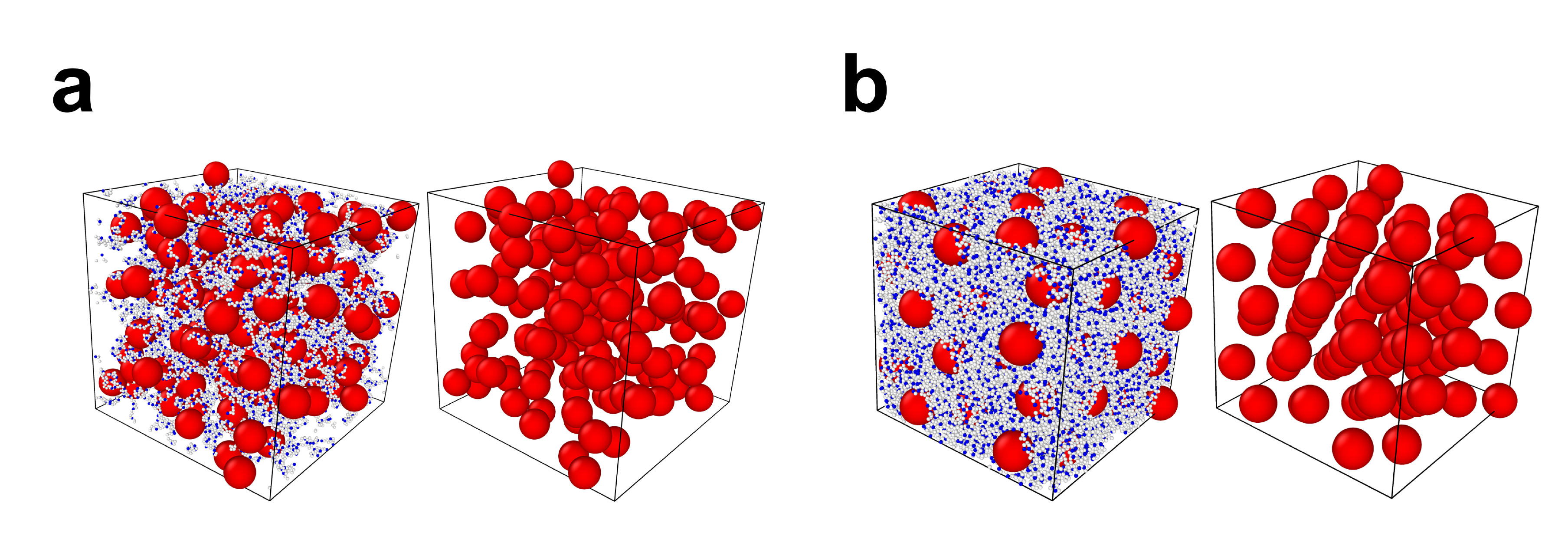}
\caption{Internal structure of hybrid coacervate formed from PEs and colloids with different charges: a) for low charge, $Q = 24e$, hybrid coacervate is liquid (disordered); b) for high charge, $Q=150e$, hybrid coacervate is the crystal of the colloids. For each figure, the left images show both polymers and colloids, while the left images show colloids only. The simulation parameters are $N = 120$ and $N=750$ for $Q = 24e$ and $Q=150e$, respectively, $f = 0.2$, and $l_{B} = \sigma$. }
\label{fig:snap-order}
\end{figure}

The results presented so far are primarily related to Regimes II and III; testing the scaling laws in Regime I is challenging. Figures~\ref{fig:phi} and \ref{fig:thickness} demonstrate that simulations of colloids with a higher charge $Q$ are required to do that. However, when the $Q$ value is too large, hybrid coacervate ceases to be liquid/amorphous. As shown in Figure~\ref{fig:snap-order}b, colloids start to exhibit long-range order and the formation of the colloidal crystal takes place.~\cite{chaikin1982} For the particular case of $f = 0.2$ and $R = 5 \sigma$, the hybrid coacervate is amorphous at $Q=60e$ but crystallizes into a colloidal crystal with long-range order at $Q = 150e$.

Colloids can be considered as bulky multivalent ions immersed in an oppositely charged, neutralizing background provided by the PEs. The crystallization of the colloids is driven by Coulomb repulsions between them and can therefore be considered as the formation of 3D Wigner (ionic) crystal. The order-disorder transition and the formation of the colloid (super)crystalline lattice are supported by the shape of the colloid structure factors shown in Figure S5a-b of the SI. For low colloid charge, the colloid structure factor exhibits a shape typical for liquids, while at high charges, the appearance of the sharp peaks reveals the long-range order. 

The Landau theory of weak crystallization suggests that the first ordered phase in a 3D Wigner crystal should be bcc.~\cite{alexander1978} To distinguish between the different packing geometries (bcc, fcc, and hcp) and specify the symmetry of the colloid lattice, the respective bond-orientational order parameters~\cite{nelson1983} have been calculated:~\cite{dellago2005} $Q_4 = 0.071 \pm 0.003$, $Q_6 = 0.428 \pm 0.002$, $W_4 = 0.003 \pm 0.004 $, and $W_6 = 0.011 \pm 0.001$. These values do not exactly coincide with those for the perfect bcc lattice, which are equal to $Q_4^{bcc} = 0.082$, $Q_6^{bcc} = 0.501$, $\hat{W}_4 = 0.159$, and $\hat{W}_6 = 0.013$. At the same time, they also do not match the perfect fcc or hcp lattice. We cannot therefore accurately determine the type of the crystalline order, and primarily attribute that to the finite-size effects, i.e., the small size of the colloidal (super)crystal comprising only 640 colloids. Such small systems may also exhibit icosahedral packing without long-range translational order.~\cite{nelson1983} Another reason may be the slow relaxation within the crystalline phase.

However, the non-zero values of the order parameters for coacervates comprising highly charged colloids, $Q = 150e$, provides the confirmation of the long-range order emergence accompanying crystallization. At low colloid charge (equal to $24e$), negligible values of the order parameters are observed in simulations, e.g., $Q_4 = 0.011 \pm 0.002$ and $Q_6 = 0.022 \pm 0.003$, are consistent with a disordered, liquid-like structure in a hybrid coacervate, which is shown in Figure~\ref{fig:snap-order}a.

\section{Bulk Modulus of Colloid-Polyelectrolyte Coacervates}
\label{section:modulus}

Based on the predictions for the internal structure of hybrid coacervates, the scaling theory was further developed to predict the osmotic compressibility of these phases and its dependence on the properties of the PE and colloid.~\cite{artem2022hybrid} In what follows, we focus on how the bulk (osmotic) modulus of the hybrid coacervate, which is the measure of the (osmotic) compressibility under uniform external pressure along all three dimensions, is governed by the colloid radius and charge. According to our scaling analysis, in the regimes of strong adsorption I and II, the osmotic modulus is independent of $Q$ and $R$ and is only controlled by the polarity of the solvent and the content of ionic monomers in the PE: 
\begin{equation}
K_{I} \simeq K_{II} \simeq uf^{2}
\label{K-I,II}
\end{equation}
This result is consistent with the scaling picture of the strongly adsorbed PE layer, which contains many layers of densely packed adsorption blobs, with the blob size increasing and the polymer density decreasing from the center to the periphery. The size of the blobs in the outermost layer is equal to that of the electrostatic blob, $\xi_{e} \simeq \left( u f^2 \right)^{-1/3}$. Osmotic properties of the hybrid coacervate are defined by the structure of the outermost layers of the colloid-PE electroneutral cells because they are in contact with each other. Using the $k_B T$ per blob rule and expressing energies in thermal units, one arrives at the result of eq.~\ref{K-I,II} $K_{I} \simeq K_{II} \simeq \xi_{e}^{-3}$. The bulk modulus is independent of the colloid properties because the colloid charge is almost entirely screened by the dense PE layer. In contrast, in Regime III, the adsorbed layer is very sparse, and interactions between the adjacent colloids are barely screened by the PE. The energy of their repulsions, $E_{Coul} \simeq u Q^2 / R$, defines the energy scale of the problem in this regime. Combining this with the typical distance $R+H_{III} \approx R$ between the colloids, one can estimate the osmotic modulus as the ratio $E_{Coul} / R^3$ and arrive at~\cite{artem2022hybrid}  
\begin{equation}
K_{III} \simeq uQ^{2}R^{-4}
\label{K-III}
\end{equation}

\begin{figure}[ht]
\centering
\includegraphics[width=0.8\linewidth]{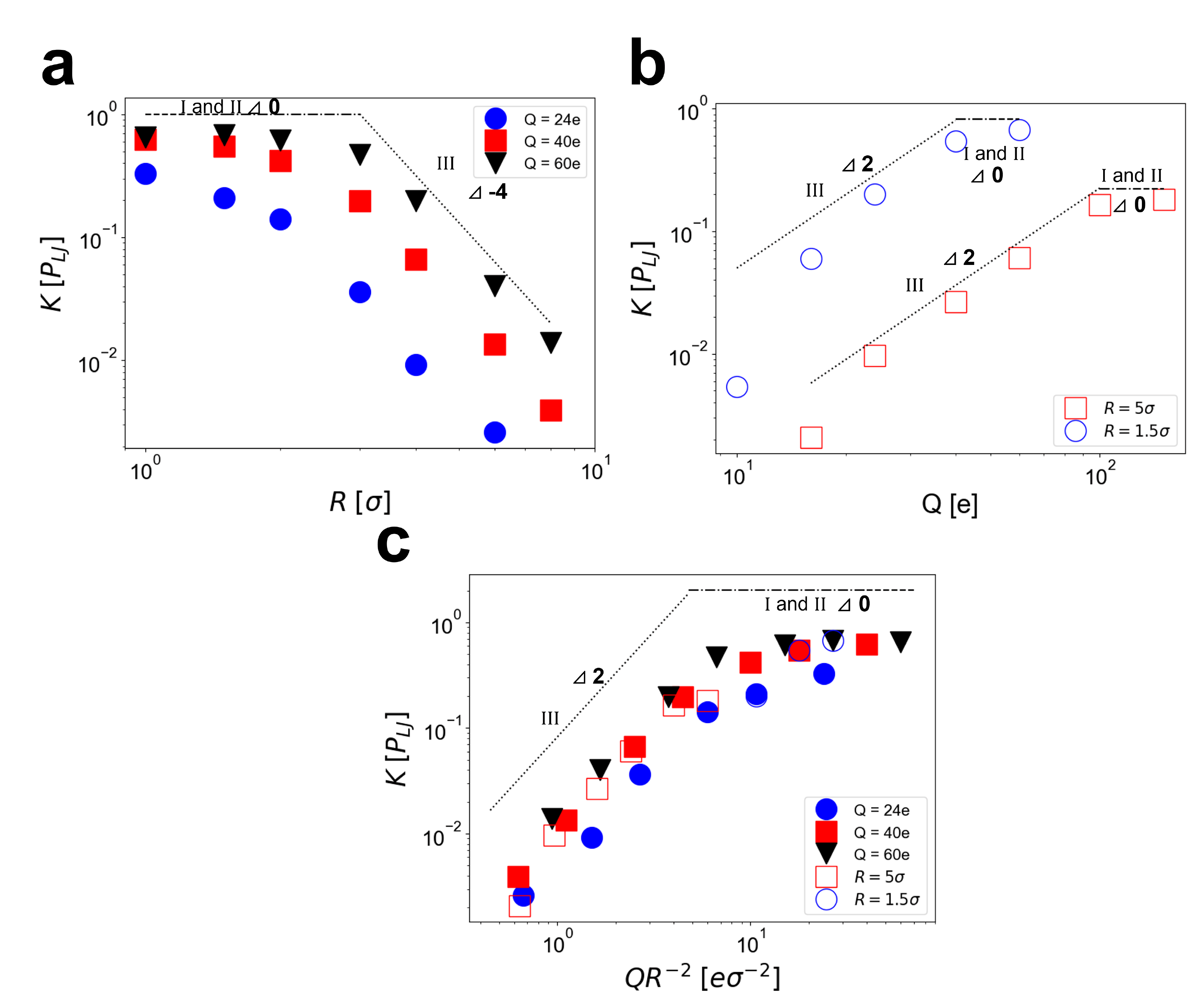}
\caption{a) The bulk modulus of the hybrid coacervate phase, $K$, as a function of 
a) the particle radius $R$ for $Q = 24e$, $40e$, and $60e$;
b) the particle charge $Q$ for $R = 1.5\sigma$ and $5\sigma$;
c) the surface charge density of the colloid, $QR^{-2}$.
In the log-log plot, theoretical scaling power laws are shown with straight lines. Colloid-PE pairs are charge-matched, $Q/e = fN$, and the simulation parameters are $f = 0.2$ and $l_{B} = \sigma$.}
\label{fig:bulk}
\end{figure}

In our simulations, the bulk modulus $K$ is calculated following the procedure of ref.~\citenum{wu2019bulk}. Detailed descriptions can be found in the Supporting Information. The dependence of $K$ and $R$ is shown in Figure~\ref{fig:bulk}a for $Q = 24e$, $40e$, and $60e$ where $P_{LJ} = \varepsilon\sigma^{-3}$ is the unit of pressure. One can see that the slope approaches $0$ in the limit of low $R$, in agreement with the scaling law for Regimes I and II (eq.~\ref{K-I,II}). For large colloids, the bulk modulus decreases with the radius $R$ and the apparent slope observed in simulations is close to the theoretical value of $-4$ given by eq.~\ref{K-III}. Similar to the results of Section~\ref{section:structure} and in agreement with eq.~\ref{crossover-II/III}, the position of the crossover between Regimes II and III shifts to larger $R$ as the colloid charge $Q$ increases.

Figure~\ref{fig:bulk}b shows that the bulk modulus is an increasing function of the colloid charge. Over a wide range of $Q$, which corresponds to Regime III, the slope observed in simulations is close to $2$. The deviation from this law only occurs at low $Q$ values, when electrostatic interactions are very weak. At high charges, this dependence plateaus, which manifests itself on the crossover to Regime II and further to Regime I. In agreement with eq.~\ref{crossover-II/III}, the higher the colloid radius, the higher the crossover position, $Q_{III/II}$.

In terms of the surface charge density, the bulk modulus can be written as $K_{III} \simeq u \left( Q / R^2 \right)^{2}$ for Regime III, while in Regimes I and II $K_{I} \simeq K_{II} \simeq u f^2 \left( Q / R^2 \right)^{0} $. These slopes, $2$ and $0$, are seen in Figure~\ref{fig:bulk}c if the data are plotted in coordinates of the charge density, albeit the collapse of all the data onto the master curve is not as good as for the layer density, $\phi$.

\newpage
\section{Mobility of Colloid Nanoparticles}
\label{section:dynamics}

It is anticipated that the coacervate dynamics strongly depend on $N$.~\cite{artem2022hybrid} For this reason, to study the particle dynamics within the hybrid coacervate phase, we release the charge matching-constraint and consider longer PEs with $fN \geq Q/e$. The charge stoichiometry between PEs and colloids is maintained to provide the global charge neutrality of the coacervate phase in the absence of counterions. We consider two cases of the colloid net charge, $Q = 24e$ and $Q = 40e$, and focus on how the diffusion of colloids changes as the function of their radius, $R$, and the chain length of the oppositely charged PE, $N$.

Ref.~\citenum{artem2022hybrid} provides scaling predictions for the dynamics of hybrid coacervates only for Region I, where most of the hybrid coacervate volume is occupied by PEs and its viscoelastic behavior is polymer-controlled. Theory suggests that, in the absence of electrostatic activation barriers induced by adsorption/desorption, the dynamics of the colloids are analogous to that of non-sticky particles in the semidilute solution of the neutral polymer.~\cite{cai2011mobility} The diffusion of the PEs can be described by Zimm-Rouse and Zimm-reptation models, and the crossover from unentangled to entangled polymer dynamics takes place as the chain length $N$ increases.~\cite{yu2020crossover,rubinstein2003polymer} In hybrid coacervates, particles smaller than the reptation tube size are not affected by the topological entanglements formed by polymer chains. Their dynamics are also Rouse-like, and the diffusion coefficient $D_{p}^{unent}$ is determined by the effective viscosity experienced by the particles.~\cite{gennes2000mobility} The latter is the Rouse viscosity of the PE chain fragments with a size comparable to the particle radius, $R$.~\cite{gennes2000mobility, cai2011mobility, schweizer2014, ge2017nanoparticle, ge2023scaling} This results in the following scaling law for small, unentangled colloids:~\cite{artem2022hybrid} 
\begin{equation}
\label{eq:D-Rouse}
D_{p, I}^{unent} \simeq \frac{k_{B}T}{\eta_{s}\phi_{I}^{2}R^{3}} \simeq D_{0}u^{−6/5}f^{−8/5}Q^{−4/5}R^{−3}
\end{equation}
where $\eta_{s}$ is the solvent viscosity and $D_{0}$ is the diffusion coefficient of a single disjointed monomer (statistical segment). In the opposite scenario, when the particle size exceeds the tube diameter, the diffusion of the colloids is constrained by topological entanglements.~\cite{cai2011mobility,ge2017nanoparticle,ge2023scaling} Their diffusion coefficient is inversely proportional to the viscosity of the entangled semidilute solution of PEs, $\eta_{rep}^{I} \simeq \eta_{s} \phi_{I}^{14/3} N^3 / N_e^2$, and reads~\cite{artem2022hybrid} 
\begin{equation}
\label{eq:D-rep}
D_{p, I}^{ent} \simeq \frac{k_B T}{\eta_{rep} R}  
\simeq D_{0}u^{−14/5}f^{−56/15}Q^{−28/15}R^{−1}N_{e}^{2}N^{-3}
\end{equation}
Here $N_{e}$ is the entanglement strand length in the melt.

As demonstrated in Sections~\ref{section:structure} and \ref{section:modulus}, the access to Regime I in our simulations is limited as it is never observed in a broad range of parameters. In addition, we are using an implicit solvent model, which does not properly reproduce hydrodynamic interactions (Zimm dynamics), which are taken into account by eqs.~\ref{eq:D-Rouse} and \ref{eq:D-rep}. This limits direct tests of the $R$ and $Q$ exponents in our dynamical scaling laws. For this reason, we primarily focus on the more universal features such as the effect of chain length, $N$, and the general applicability of the non-sticky model of the colloid diffusion within the hybrid coacervates. These aspects are independent of whether the solvent is treated explicitly or implicitly in the coarse-grained simulations.~\cite{yu2020crossover} Two particular cases of the nanoparticle, with $Q = 24e$ and $Q = 40e$, are considered.

\begin{figure}[ht]
\centering
\includegraphics[width=0.85\linewidth]{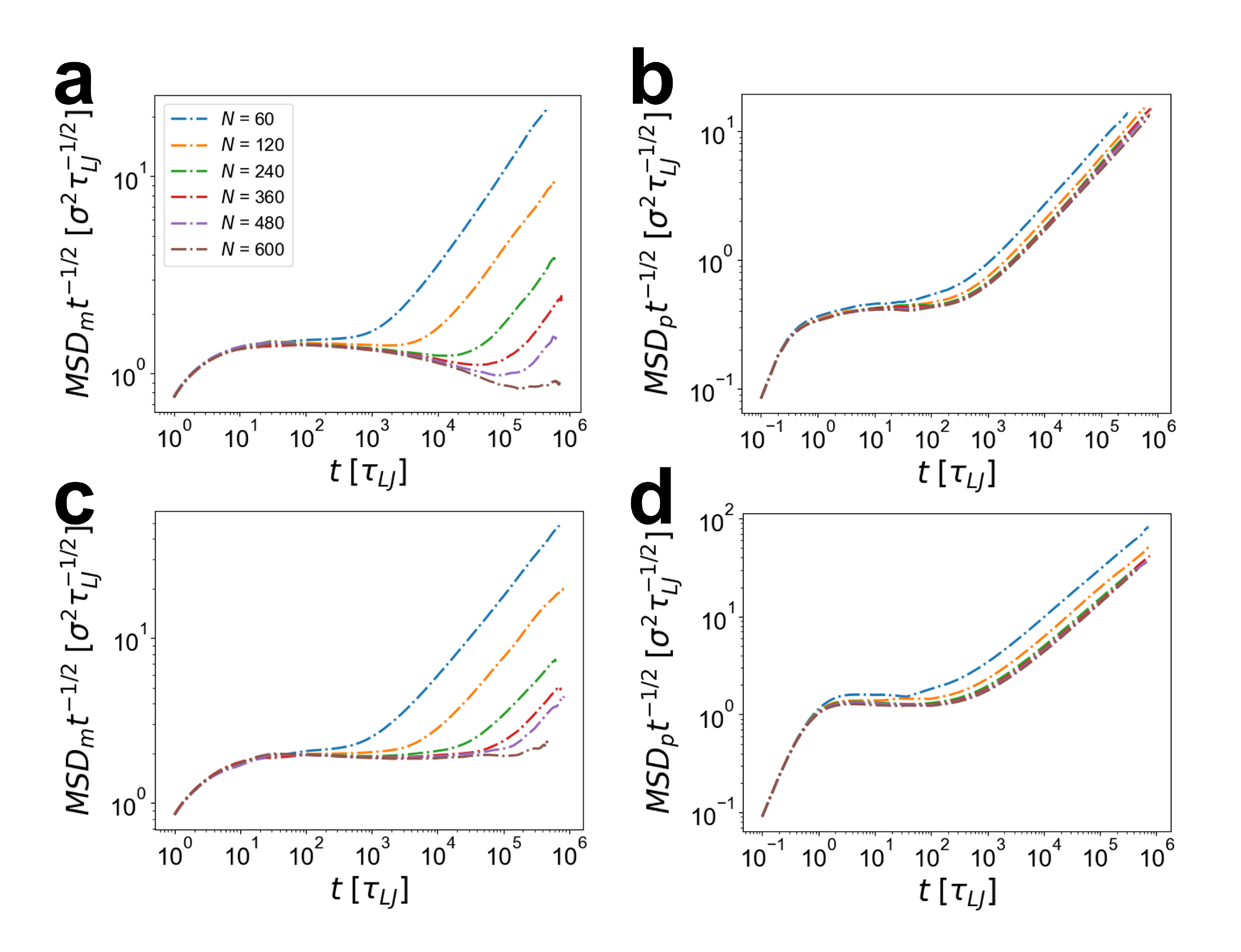}
\caption{Mean-squared displacement (MSD) of central five monomers of each PE chain $MSD_{m}$ (a,c) and of particle center of mass $MSD_{p}$ (b,d) as the function of time, $t$, for different length of the PEs, $N$. MSD is normalized by $t^{1/2}$ to easier distinguish between Rouse and reptation dynamics. Plots a) and b) correspond to $R = 2 \sigma$, and plots c) and d) to $R = 5 \sigma$.  The simulation parameters are $f = 0.2$, $Q = 24e$ and $l_{B} = \sigma$.}
\label{fig:dynamicsq24}
\end{figure}

The mean-squared displacement (MSD) of the PE monomers and of the colloid with $Q = 24e$ and $R = 2 \sigma$ is shown in Figure~\ref{fig:dynamicsq24}a-b. MSD values are normalized by the factor of $t^{-1/2}$, which makes them flat for the Rouse subdiffusion where $MSD_{m} \sim t^{1/2}$ and decreasing for one of the regimes of reptation subdiffusion where $MSD_{m} \sim t^{1/4}$. As discussed in Section~\ref{section:structure}, the selected parameters of colloids correspond to the border scaling Regime I, close to the I/II crossover. Figure~\ref{fig:dynamicsq24}a demonstrates that, as the PE chain length increases, the polymer dynamics crossovers from unentangled to entangled regime. This is indicated by the appearance of the region where $MSD_{m} t^{-1/2}$ goes down for long PEs, $N > 240$. This dynamical crossover is analogous to that observed for conventional interpolyelectrolyte coacervates and neutral semidilute solutions.~\cite{yu2020crossover}
While PE chains undergo crossover to entangled dynamics, the dynamics of colloids remain Rouse-like because the normalized MSD exhibits a plateau at intermediate time scales. This is consistent with the assumption of ref.~\citenum{artem2022hybrid} suggesting the non-sticky behavior of colloids within the hybrid coacervates and considering them as the quasi-neutral particles diffusing in the quasi-neutral polymer solution.~\cite{cai2011mobility}  

The virtual independence of the particle MSD (and therefore the diffusion coefficient) on the PE chain length is consistent with eq.~\ref{eq:D-Rouse}, derived for small particles not affected by entanglements between polymers. For the particle to \textit{feel} the entanglements, its size should be comparable to the reptation tube size, $a$. For the semidilute solution, the latter can be estimated as $a \simeq \left( b l N_{e} \right)^{1/2}\phi^{-2/3}$. Using the Kuhn length $b = 1.82\sigma$, the bond length $l = 0.96\sigma$, $N_{e} = 70$ for melts,~\cite{everaers2004rheology} and $\phi = 0.27$, one arrives at $a \approx 27 \sigma$, which by far exceeds the size of the colloid considered in simulations, $R \simeq 2 \sigma$. To consider the case of entangled colloids and test eq.~\ref{eq:D-rep} in simulations, one should use much larger $R$ values. However, even in this case, increasing the colloid radius would lead to weaker Coulomb attractions with the PEs, a lower density of the hybrid coacervate, and an even larger diameter of the reptation tube. Thus, different model parameters, e.g., chain stiffness, are required to reproduce this behavior, and this goal is beyond the scope of the present study.

Analogous simulations were also performed for larger particles with $R = 5\sigma$, which correspond to the scaling Regime III of the hybrid coacervate. Due to the much lower density of the coacervate phase, the dynamics of polymer chains remain Rouse-like even for the highest length considered, $N = 600$, as seen in Figure~\ref{fig:dynamicsq24}c. Colloid nanoparticles also demonstrate Rouse subdiffusion, as demonstrated in Figure~\ref{fig:dynamicsq24}d. 

\begin{figure}[ht]
\centering
\includegraphics[width=0.4\linewidth]{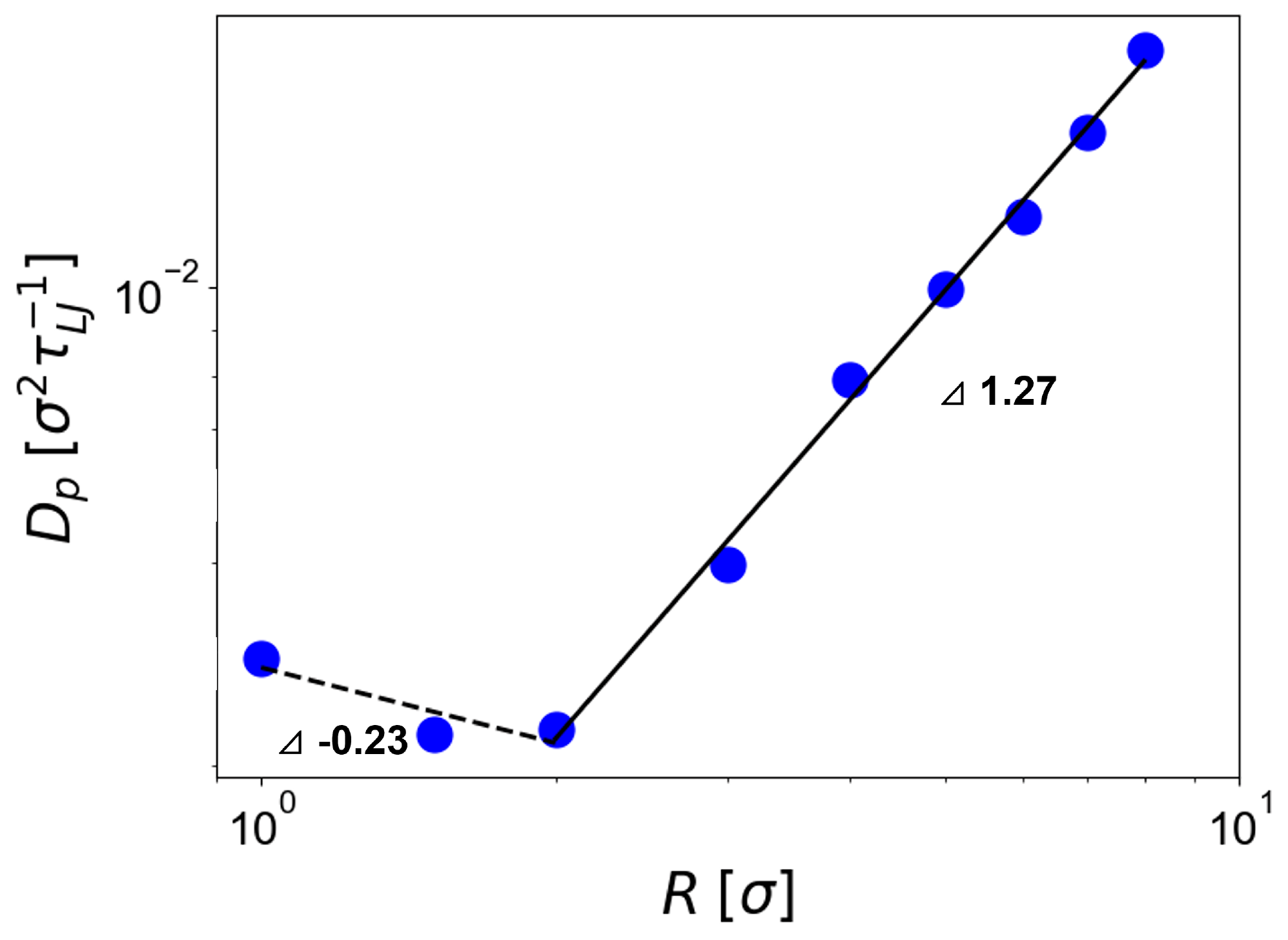}
\caption{The dependence of the colloid diffusion coefficient $D_{p}$ on the colloid radius $R$. Straight lines provide a  fit of the simulation data with the power law in the regions of decreasing and increasing $D_{p}$; the slopes are equal to $-0.23$ and $1.27$. The errors were estimated by dividing the particles into five groups and calculating the standard deviation of the diffusion coefficient for each group. The errors are within the size of each data point. The simulation parameters are $N = 120, f = 0.2$, $Q = 24e$, and $l_{B} = \sigma$.}
\label{fig:D_vs_R}
\end{figure}

Rouse-like diffusion of colloids for both $R = 2 \sigma$ and $R = 5 \sigma$ indicates that a similar theoretical approach can be potentially applied in Regimes II and III to describe their diffusion albeit the viscosity of the hybrid coacervate may be already not entirely by polymers, as discussed in ref.~\citenum{artem2022hybrid}. If so, the diffusion coefficient of small unentangled colloids should obey
\begin{equation}
D_{p, II}^{unent} \simeq \frac{k_{B}T}{\eta_{s}\phi_{II}^{2}R^{3}} \sim R^{-1/3}
\label{eq:D-Rouse-II}
\end{equation}
\begin{equation}
D_{p, III}^{unent} \simeq \frac{k_{B}T}{\eta_{s}\phi_{III}^{2}R^{3}} \sim R^{7/3}
\label{eq:D-Rouse-III}
\end{equation}
for Regimes II and III, respectively. Comparing these results to eq.~\ref{eq:D-Rouse} one can conclude that an increase in the colloid radius, which triggers a continuous system evolution from Regime I through Regime II to Regime III, should be accompanied by a non-monotonic change of the colloid mobility. The latter first decreases with the colloid size in Regimes I and II, but then increases in Regime III because of the strong drop in the PE layer density. Interestingly, this non-monotonic trend is indeed detected in our simulations. The respective dependence is shown in Figure~\ref{fig:D_vs_R}, and the diffusion coefficients of colloids were obtained from the linear fit of the MSD curves with $MSD_{p} = D_{p} t$ in the region of normal diffusion, i.e., of large times $t$. However, the apparent slopes obtained from Figure~\ref{fig:D_vs_R} are different and should not be directly compared to the theoretical eqs.~\ref{eq:D-Rouse}, \ref{eq:D-Rouse-II} and \ref{eq:D-Rouse-III} because simulations with implicit solvent do not reproduce hydrodynamics, i.e., Zimm dynamics.

\begin{figure}[ht]
\centering
\includegraphics[width=0.85\linewidth]{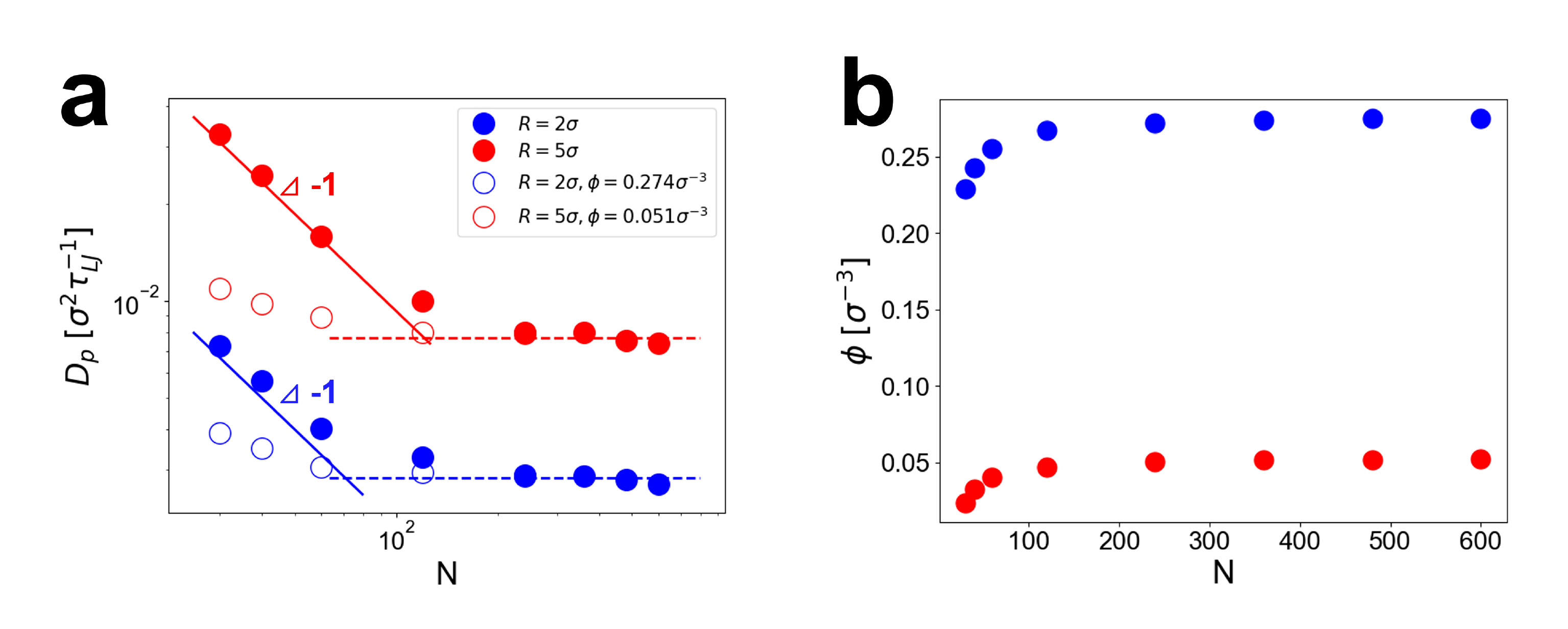}
\caption{ a) Diffusion coefficient of particles $D_{p}$ as a function of PE chain length $N$. Solid and dash lines have a slope $-1$ and $0$, respectively. The errors are within the symbol size and were estimated by dividing the particles into five groups and calculating the standard deviation of the diffusion coefficient for each group. 
b) Polymer layer density $\phi$ as a function of PE chain length $N$. The errors were estimated by the standard deviation of sampling data after equilibrium. The simulation parameters are $f = 0.2$, $Q = 24e$, and $l_{B} = \sigma$.}
\label{fig:D_vs_N}
\end{figure}

The dependence of the colloid diffusion coefficient on the PE chain length $N$ is shown in Figure~\ref{fig:D_vs_N}a. Large colloids with $R = 5 \sigma$ are more mobile as compared to their small counterparts with $R = 2 \sigma$ because the PE density of the hybrid coacervate in the former case is approximately 5 times lower, see Figure~\ref{fig:D_vs_N}b. For both particle sizes, the particle diffusion coefficient first decreases with the increase of $N$ and then reaches a plateau and become $N$-independent. The constant value of the diffusion coefficient for long PEs is consistent with the scaling law given by eq.~\ref{eq:D-Rouse}. This behavior is violated for short chains when $D$ significantly decreases with increasing $N$. The reason for that may be two-fold. Eq.~\ref{eq:D-Rouse} was derived under the assumptions of (i) $N$-independent density of the hybrid coacervate phase and (ii) the PE end-to-end distance substantially exceeds the colloid radius.~\cite{artem2022hybrid} 

If condition (i) is fulfilled but (ii) is violated, i.e. the chains are smaller than the colloid, then the colloid experiences an effective viscosity $\eta_{R} \simeq \eta_{s} \phi^{2} N$ equal to the Rouse viscosity of the semidilute solution. This leads to the diffusion coefficient decreasing as $D_{p} \simeq k_B T / \eta_{R} R \sim N^{-1} $ at increasing polymer length,~\cite{cai2011mobility} which is similar to the decrease seen in Figure~\ref{fig:D_vs_N}a. However, the increasing density of the hybrid coacervate at low $N$, which is shown in Figure~\ref{fig:D_vs_N}b, also contributes to the decrease in $D$. To factorize the density and the chain size effects, (i) and (ii), the NVT simulations were performed for the hybrid coacervates comprising short PEs, but with the fixed density equal to that in the limit of high $N$. The resulting diffusion coefficient values are shown in Figure~\ref{fig:D_vs_N}a with the open symbols. After the density adjustment, the decrease in $D_{p}$ with increasing $N$ in the range of short chains is very weak, suggesting that the density effect is dominant, while the PE length effect is much weaker. This is because even for the shortest PEs studied, $N=30$, their size $R_{e} \simeq \left( l b N \right)^{1/2} \simeq 7.2 $ is comparable or larger than the radius of the colloids, $R  = 5 \sigma$ and $R = 2 \sigma$.

\begin{figure}[ht]
\centering
\includegraphics[width=0.85\linewidth]{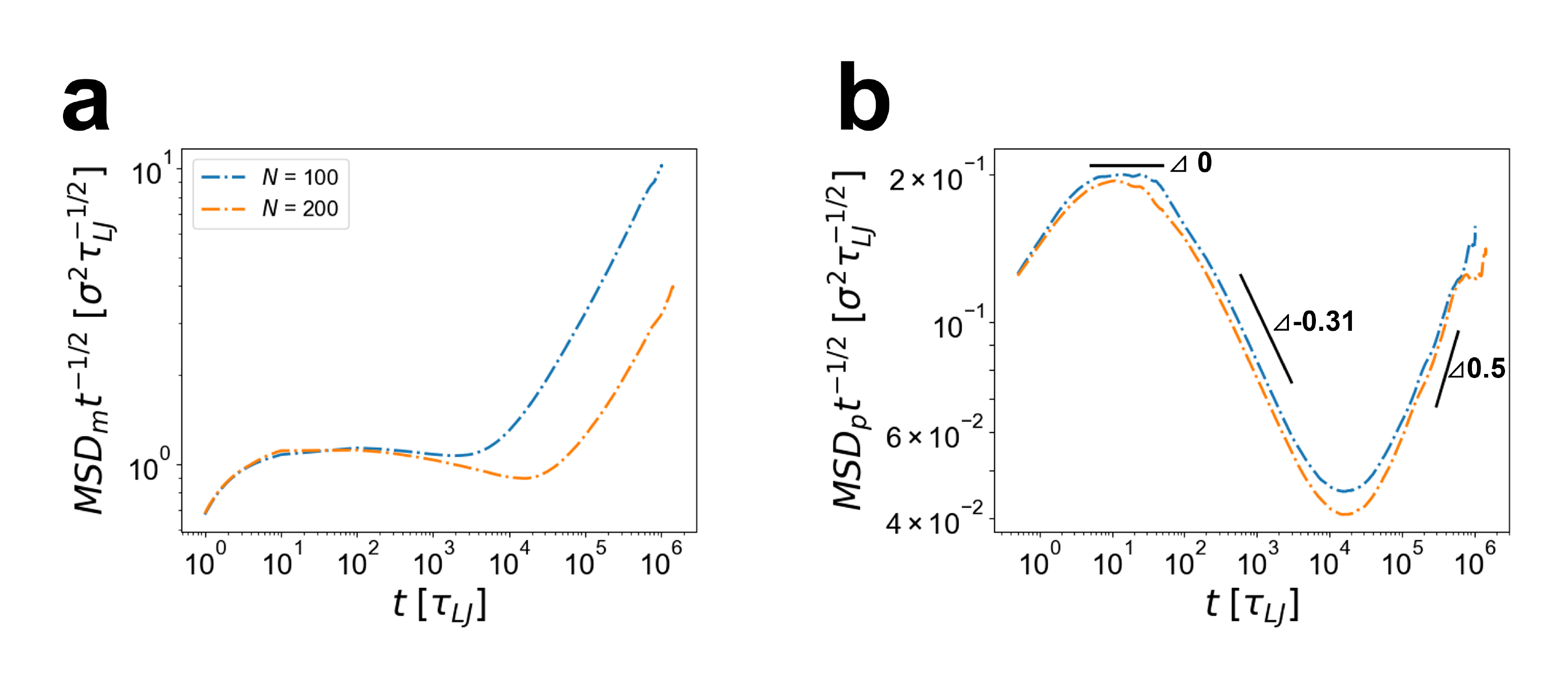}
\caption{
a) MSD of central five monomers of each PE chain $MSD_{m}$ normalized by the factor of $t^{-1/2}$ as a function of simulation time $t$ for hybrid coacervates formed from colloids with the radius $R = 2\sigma$ and PEs of the length  $N=100$ or $N = 200$.  
b) MSD of colloids $MSD_{p}$ normalized by the factor of $t^{1/2}$ as a function of $t$ in the same systems. The simulation parameters are $f = 0.2$, $Q = 40e$, and $l_{B} = \sigma$.}
\label{fig:largeQ}
\end{figure}

Our results indicate that, for colloids of sufficiently low charge ($Q = 24e$ in this particular case), their dynamics can be reasonably described by classical non-sticky models. Moreover, simulations show that this approach may be applicable not only to Regime I, as suggested in ref.~\citenum{artem2022hybrid}, but also in Regimes II and III. However, as the colloid charge increases, the electrostatic attractions between PE and colloids are so strong that the activation energy for desorption exceeds the thermal energy, and the particle is effectively sticky.~\cite{artem2022hybrid} The normalized MSDs of the PE monomers and the colloid for $Q = 40e$ and $R = 2 \sigma$ are shown in Figure~\ref{fig:largeQ}b. In contrast to the case for a lower charge, these colloids demonstrate subdiffusion, which is slower than Rouse-like, as inferred from the decreasing behavior of $MSD_{m} / t^{1/2}$ at intermediate time scales. The small size of the nanoparticles suggests that they should not be directly affected by topological entanglements between PEs, with are noticeable for $N = 200$, as seen in Figure~\ref{fig:largeQ}a. For this reason, we conclude that the slow, non-Rouse-like colloid dynamics should be attributed to an adsorption/desorption-related mechanism. The respective electrostatic activation barriers make colloid-PE interactions effectively sticky, and the underlying physics, which controls the mobility of highly charged colloids, is more complicated.

Here we refer to recent work~\cite{yamamoto2018theory} that studied the dynamics of sticky neutral colloids in melts of neutral polymers, which also reported $D (N)$ dependencies, which decrease at low $N$ and plateau for high $N$. This is similar to the behavior of the density-adjusted diffusion coefficient shown in Figure~\ref{fig:D_vs_N}. Ref.~\citenum{yamamoto2018theory} attributed this behavior to the colloid mobility crossover from the ``core-shell'' mechanism to the ``vehicle'' mechanism. The former suggests that the colloid is surrounded by the polymer shell formed by the adsorbed layer and the friction is effectively experienced by this core-shell particle, whose radius exceeds that of the bare colloid by the polymer chain size. As polymers become longer, colloids start to diffuse together with the fragment of the long polymer chain, which is adsorbed on them and serves as a vehicle; this process is accompanied by random activated events of chain adsorption and desorption, which provide the change of the vehicle.~\cite{yamamoto2018theory} To what extent these mechanisms can be applied to hybrid coacervates, where PE adsorption is driven by long-range Coulomb rather than short-range sticky interactions, is an open problem that will require a separate, comprehensive study combining theoretical and simulation approaches.

\section{Conclusions}
\label{section:conclusion}

In this work, systematic coarse-grained simulations of salt-free hybrid coacervates of linear polyelectrolytes and oppositely charged colloid nanoparticles have been performed in the NPT ensemble. A Kremer-Grest model supplemented by Coulomb interactions was been employed to describe the charged polymers. Colloids were modeled as hard impenetrable spheres. The structural, osmotic, and dynamic properties of hybrid coacervates were explored over a wide range of parameters, including the colloid net charge and radius. Simulation results were compared to the scaling theory of hybrid colloid-polyelectrolyte coacervates,~\cite{artem2022hybrid} and many predictions of the latter were successfully corroborated.     

For structural properties, the simulations demonstrated that increasing colloid radius $R$ or decreasing charge $Q$ triggers the evolution of the PE layer. This corresponds to the transitions from the scaling regime of strong spherical adsorption (Regime I) to the strong quasi-planar adsorption (Regime II) to the weak quasi-planar adsorption (Regime III). For the quasi-planar Regimes, II and III, the dependence of the average density of the PE layer, $\phi$, collapses onto a master curve when expressed in terms of the surface charge density, $Q/R^2$. This universality of the hybrid coacervate structure and the observed power laws for the $\phi (Q/R^2)$ dependence are in agreement with the scaling analysis.~\cite{artem2022hybrid} The thickness $H$ of the PE layer,  which separates the adjacent colloids, is found to be a non-monotonic function of the colloid radius and charge, as was theoretically anticipated, albeit these dependencies demonstrate a weaker universal behavior than the dependence of $\phi$ on $Q/R^2$.

For hybrid coacervates with very highly charged colloids, we observed the formation of a colloidal crystal, which cannot be adequately described by the existing scaling theory developed for the disordered phase only. The long-range order of the nanoparticles within the polyelectrolyte matrix is driven by their Coulomb repulsions.

To understand how the hybrid coacervates respond to uniform compression, we examined the dependence of the osmotic (bulk) modulus $K$ of the hybrid coacervate on the properties of the colloid and polyelectrolyte. At a high colloid charge and/or its small radius, when PE adsorption is strong, the osmotic compressibility of the phase is virtually independent of the colloid properties. This is consistent with the theory predicting that $K$ is controlled by the polymer for the strong adsorption regimes, I and II. In contrast, when adsorption is weak, scaling suggests that it is defined by the colloids and $K_{III} \sim (Q / R^2)^2$. This power law is reproduced in simulations of bulky particles carrying a sufficiently low charge.  

To quantify the mobility of the colloids, their mean-squared displacement (MSD) was obtained as a function of time for the hybrid coacervates comprising PEs of different lengths. For low $Q$ values, when Coulomb interactions are not very strong, the diffusion of the colloids can be adequately described by considering them as non-sticky particles. Their MSD resembles that of the monomers of unentangled polymers in semidilute solutions, with the Rouse subdiffusion at intermediate times, $MSD_{p} \sim t^{1/2}$, followed by normal diffusion in the terminal regime, $MSD_{p} \sim t$. In our simulations, the particle size was much lower than the reptation tube diameter, and their mobility was not affected by the Rouse-to-reptation crossover in the PE dynamics. The increase in the colloid diffusion coefficient is only observed for short PE chains, much below the onset of reptation, and we primarily attribute that to the low density of the coacervate phase at low $N$. After the diffusion coefficient of colloids $D_{p}$ is normalized to account for the change in the polymer density, it is barely dependent on $N$. This result supports the applicability of the non-sticky diffusion models to the colloids carrying moderate charge. Interestingly, scaling predicts the non-monotonic dependence of $D_{p}$ on $R$, which is also detected in our simulations.

However, colloid dynamics are more complex when $Q$ becomes sufficiently high. In this case, the subdiffusion at intermediate time scales is slower than the Rouse subdiffusion. The apparent slope in the $MSD_{p} \sim t^{\alpha}$ dependence is close to $\alpha \approx 0.19$. We attribute this behavior to the strong Coulomb attractions between colloids and polyelectrolytes, which generated a substantial activation barrier for the elementary adsorption/desorption processes. Therefore, colloids should be viewed as effectively sticky for polymers, and the theoretical framework for non-sticky particles is no longer applicable.~\cite{artem2022hybrid} More comprehensive theoretical and simulation efforts are required in the case of very strong Coulomb interactions between colloids and polyelectrolytes to better understand the rheology of condensed phases and colloid mobility in them.

To summarize, the simulation findings presented here are consistent with the experimental literature~\cite{cummings2018phase, yeong2020formation} and help validate the scaling theory developed in our earlier work.~\cite{artem2022hybrid} They serve as important guidelines for the rational design of hybrid materials derived from the complexation of polyelectrolytes with oppositely charged proteins, surfactants, and solid nanoparticles. In addition, simulation results point towards new and promising avenues through which the theory of hybrid coacervates could be further developed.

\section*{Supporting Information}
1. Effect of the Colloid Charge Assignment on the Coacervate Properties; 2. Radial Density Profile of Adsorbed PE Layers; 3. Radial Distribution Function of Colloids and Ionic Monomers; 
4. Bulk Modulus Calculation; 5. Structure Factors of the Colloids in the Hybrid Coacervates.

\section*{Acknowledgement}
This work was supported by the Department of Energy, Basic Energy Sciences, Division of Materials Science and Engineering.

\bibliography{bibliography.bib}

\end{document}


\renewcommand{\thefigure}{S\arabic{figure}}
\renewcommand{\thetable}{S\arabic{table}}

\newpage

\section{Effect of the Colloid Charge Assignment on the Coacervate Properties}
\label{subsection:charge-assignment}

To test the effects of charge assignment for each particle on the coacervate properties, we use the same model as described in the main text except that the charges are not assigned as a point-like charge at the center of the particle. Instead, for each particle, we discretize the spherical surface into $n$ virtual interaction sites by Fibonacci sphere algorithm,~\cite{Gonzalez2010-bg} and each site carries the charge $z_{i} = Q/n$. Then we group the spherical particle (with no charge now) and its $n$ virtual charged sites into one rigid body, i.e. the new charged particle. To maintain the same excluded volume interactions, each virtual site has no LJ interactions with other beads, particles, or sites; it only interacts with other charged beads or sites via Coulomb forces. In addition, for the new charged particle, we assign $m = 0.5$ to the spherical particle and $m = 0.5/n$ to each virtual charged site so that the new particle has the same mass as the original particle with the net charge in the center.

\begin{table}[h!]
\centering
\begin{tabular}{||c c c c||} 
 \hline
 charge distribution & $\phi$ & $H [\sigma]$ & $D_{p} [\sigma^{2}\tau_{LJ}^{-1}] \times [10^{3}\tau_{LJ}]$ \\ [0.5ex]  
 \hline\hline
 $n = 1$ & $0.267 \pm 0.001$ & $2.18 \pm 0.01$ & $3.4 \pm 0.2$ \\ 
 $n = 64$ & $0.267 \pm 0.001$ & $2.18 \pm 0.01$ & $3.3 \pm 0.2$ \\
 $n = 128$ & $0.267 \pm 0.001$ & $2.17 \pm 0.01$ & $3.5 \pm 0.2$ \\
 $n = 256$ & $0.268 \pm 0.001$ & $2.17 \pm 0.01$ & $3.5 \pm 0.2$ \\[1ex] 
 \hline
\end{tabular}
\caption{Structural and dynamic properties of the hybrid coacervate phase for the different distributions of the particle charge: $n = 1$ denotes the case of the single charge at the center of the particle; $n = 64$, $128$, and $256$ represent different degrees of the sufficiently uniform charge smearing throughout the particle surface. $\phi$ is the polymer layer density, $H$ is the polymer layer thickness, and $D_{p}$ is the diffusion coefficient of particle center of mass.
The simulation parameters are $N = 120$, $f = 0.2$, $R = 2 \sigma$, $Q = 24e$, and $l_{B} = \sigma$. 
}
\label{table:2}
\end{table}

\begin{table}[h!]
\centering
\begin{tabular}{||c c c c||} 
 \hline
 charge distribution & $\phi$ & $H [\sigma]$ & $D_{p} [\sigma^{2}\tau_{LJ}^{-1}] \times [10^{5}\tau_{LJ}]$ \\ [0.5ex]  
 \hline\hline
 $n = 1$ & $0.411 \pm 0.001$ & $2.45 \pm 0.01$ & $2.0 \pm 0.1$ \\ 
 $n = 128$ & $0.411 \pm 0.001$ & $2.46 \pm 0.01$ & $1.9 \pm 0.1$ \\
 $n = 256$ & $0.410 \pm 0.001$ & $2.45 \pm 0.01$ & $2.0 \pm 0.1$ \\[1ex] 
 \hline
\end{tabular}
\caption{Structural and dynamic properties of hybrid coacervate phase for the different particle charge distributions: $n = 1$ corresponds to the charges at the center of the particle; $n = 128$ and $256$ represent different degrees of the charge smearing throughout the particle surface. $\phi$ is the polymer layer density, $H$ is the polymer layer thickness, and $D_{p}$ is the diffusion coefficient of the particle center of mass (i.e. the center of the partcile).
The simulation parameters are $N = 200$, $f = 0.2$, $R = 2\sigma$, $Q = 40e$, and $l_{B} = \sigma$.
}
\label{table:3}
\end{table}

We perform the same NPT simulations to study the coacervate phase formed from the PE chains and the charged particles with different charge distributions, i.e. different $n$ values. The structural and dynamic properties of the hybrid coacervate phases for different charge distributions (i.e., different $n$ values) are listed in Table~\ref{table:2} and~\ref{table:3}. One can see that the coacervate properties for the particles with the charges assigned at their center ($n = 1$) are the same as for the particles carrying the surface charges with various degrees of surface charge smearing ($n = 64$, $128$, and $256$).

\begin{figure}
\centering
\includegraphics[width=0.65\linewidth]{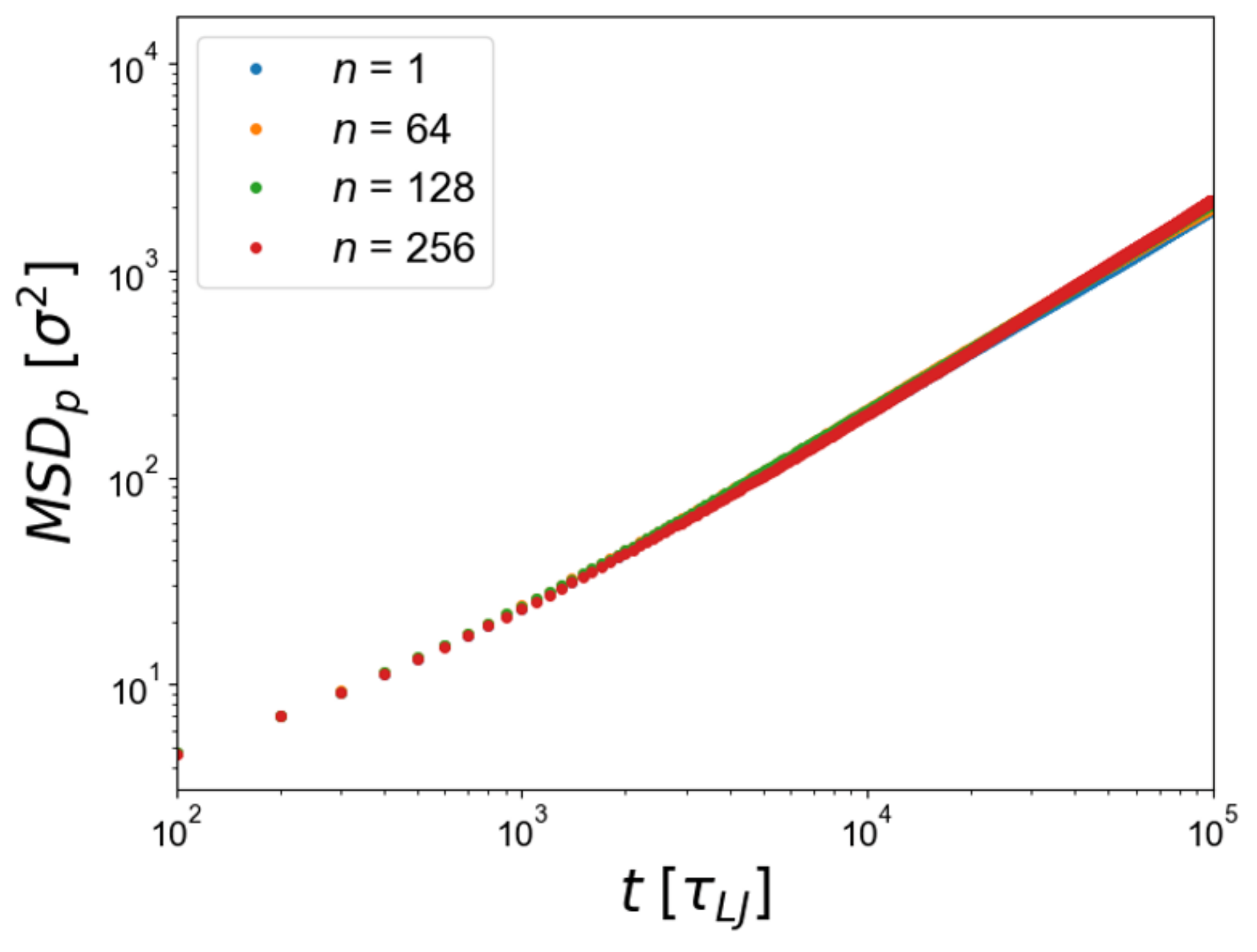}
\caption{Mean squared displacement (MSD) of the center of mass of the charged particles, $MSD_{p}$, as the function of time (log-log plot). $n = 1$ is the particle with the charge at its center; $n = 64$, $128$, and $256$ represent different degrees of the charge smearing. The simulation parameters are $N = 120$, $f = 0.2$, $R = 1/5\sigma$, $Q = 24e$, and $l_{B} = \sigma$.}
\label{fig:MSD-SI}
\end{figure}

\begin{figure}[ht]
\centering
\includegraphics[width=\linewidth]{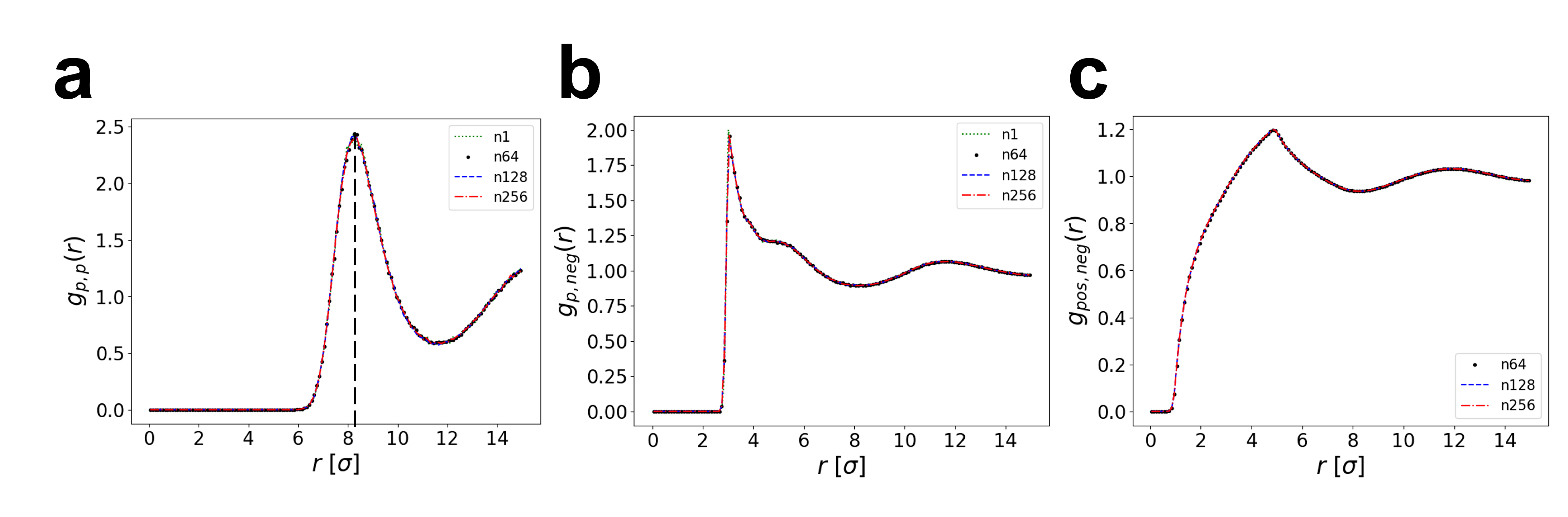}
\caption{
a) Radial distribution function (RDF) of the particle, $g_{p, p}(r)$, as the function of the distance between their center of mass, $r$,  The black dash line shows the position of the first peak. 
b) RDF for the particle center of mass and the negatively charged monomer $g_{p, neg}(r)$.  
c) RDF for the positively charged sites on the particle surface and the negatively charged polyelectrolyte monomers, $g_{pos, neg}(r)$.
For all the plots, the simulation parameters are $N = 120$, $f = 0.2$, $R = 2\sigma$, $Q = 24e$, and $l_{B} = \sigma$. Different curves correspond to the different charge distributions on the colloid particle: $n1$ corresponds to the particle with a single point-like charge in the center; $n64$, $n128$, and $n256$ represent particles with the charges evenly distributed at the particle's surface (see Section~\ref{subsection:charge-assignment} for details).
}
\label{fig:rdf}
\end{figure}

We can further compare their internal structures by calculating the partial radial distribution functions (RDF) between the different species in the coacervate phase. Figure~\ref{fig:rdf} in Section~\ref{subsection:rdf} shows that the curves for the different degrees of the surface charge smearing, $n = 64$, $128$, and $256$, and particle with the point-like charge, $n=1$, perfectly coincide.

Finally, in Figure~\ref{fig:MSD-SI}, we construct the mean squared displacement (MSD) of the center of mass of the charged particles as the function of time to reveal both the short- and the long-time diffusion behavior of the original and the new type of particles. It is seen that the distribution of charges changes the particle diffusion at neither short nor long time scales.

\newpage

\section{Radial Density Profile of Adsorbed PE Layers}
\label{subsection:distribution}

\begin{figure}
\centering
\includegraphics[width=0.6\linewidth]{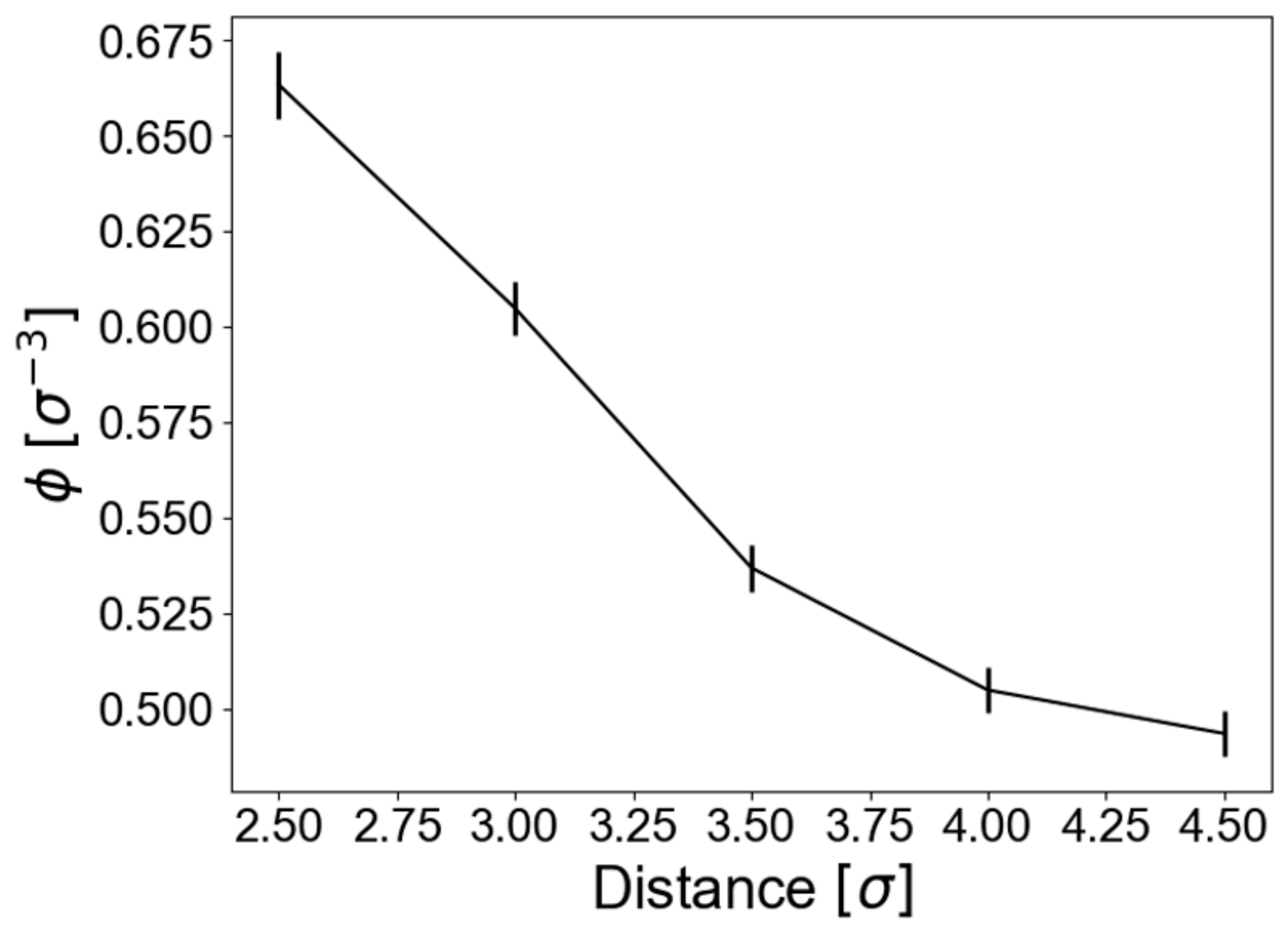}
\caption{The layer density $\phi$ as a function of the distance from the particle center. The simulation parameters are $N = 200$, $f = 0.2$, $R = 1.5\sigma$, $ Q = 40$, and $l_{B} = \sigma$.}
\label{fig:distri}
\end{figure}

\newpage

\section{Radial Distribution Function of Colloids and Ionic Monomers}
\label{subsection:rdf}

\newpage

\section{Bulk Modulus Calculation}
\label{subsection:bulk}

\begin{figure}[ht]
\centering
\includegraphics[width=0.65\linewidth]{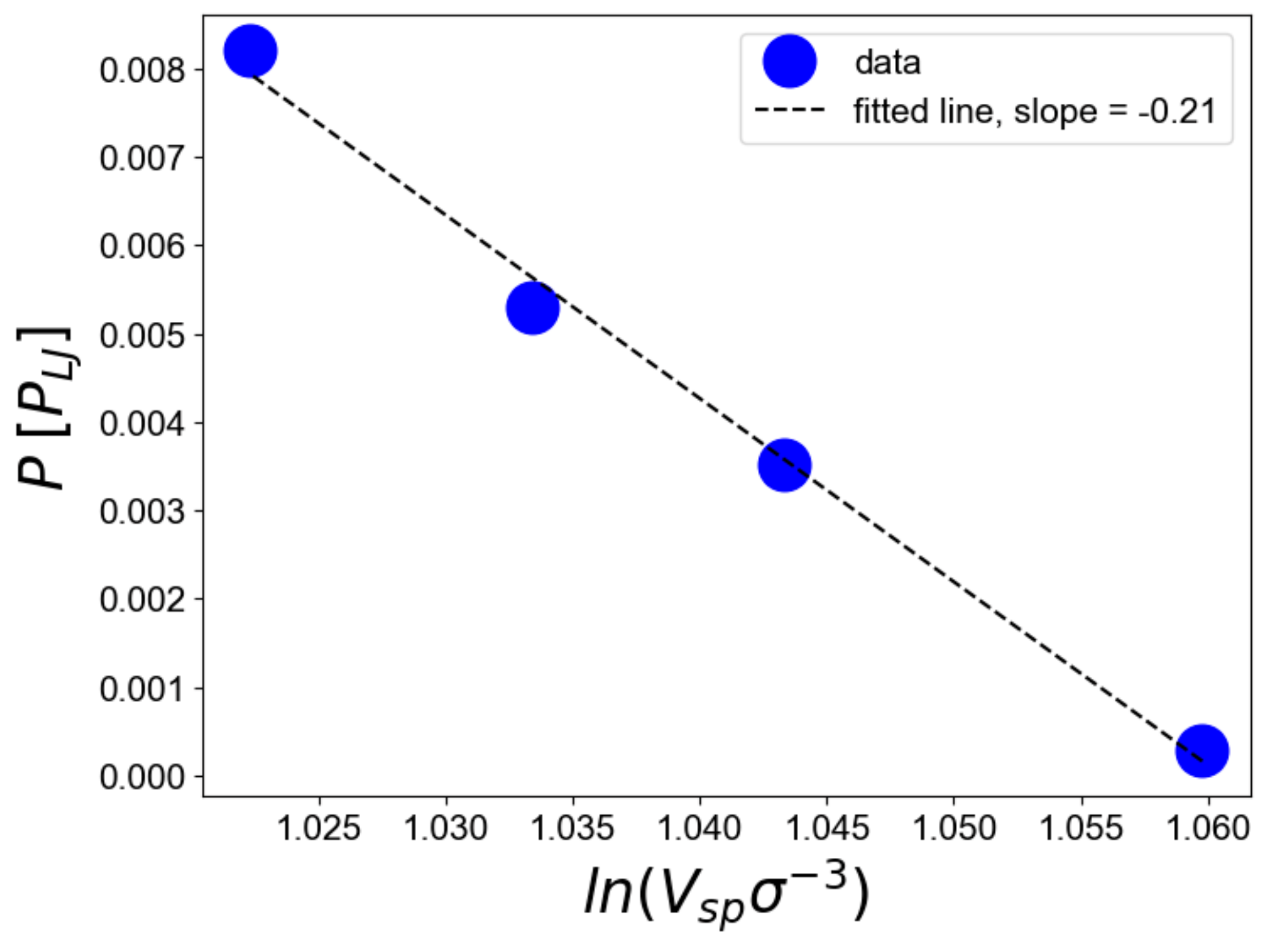}
\caption{The relationship between the applied external pressure $P$ and the natural logarithm of the specific volume $\ln V_{sp}$ per interacting entity (monomer bead or the particle). Simulation data are shown in blue points. A linear function fits data, and the legend shows the slope. The simulation parameters are $N = 120$, $f = 0.2$, $R = 1.5\sigma$, $Q = 24e$, and $l_{B} = \sigma$.}
\label{fig:bulk-fit}
\end{figure}

The bulk (or the osmotic) modulus $K$ of the hybrid coacervate phase is calculated in the following way.~\cite{wu2019bulk} First, an external pressure $P$ is applied to compress the coacervate phase uniformly along three orthogonal directions, and the corresponding volume $V$ is measured. Then this process is repeated for different $P$ values to obtain the relationship between $P$ and the specific volume $V_{sp} = V/N_{total}$ per bead of any type. Here $N_{total}$ is the total number of the polymer monomers and particles in the system; recall that it is assumed that all beads have the same unit mass. To maintain the linear relationship assumption, the volume change is limited to the maximal of $5\%$.~\cite{wu2019bulk} Finally, the bulk modulus is obtained from the slope of the line fitting the dependence of the pressure on the natural logarithm of the specific volume: $K = -{dP} / {d(\ln V_{sp})}$. The representative example of this fitting procedure is shown in Figure~\ref{fig:bulk-fit}.

\newpage
\section{Structure Factors of the Colloids in the Hybrid Coacervates}
\label{subsection:sq-SI}

\begin{figure}[ht]
\centering
\includegraphics[width=0.8\linewidth]{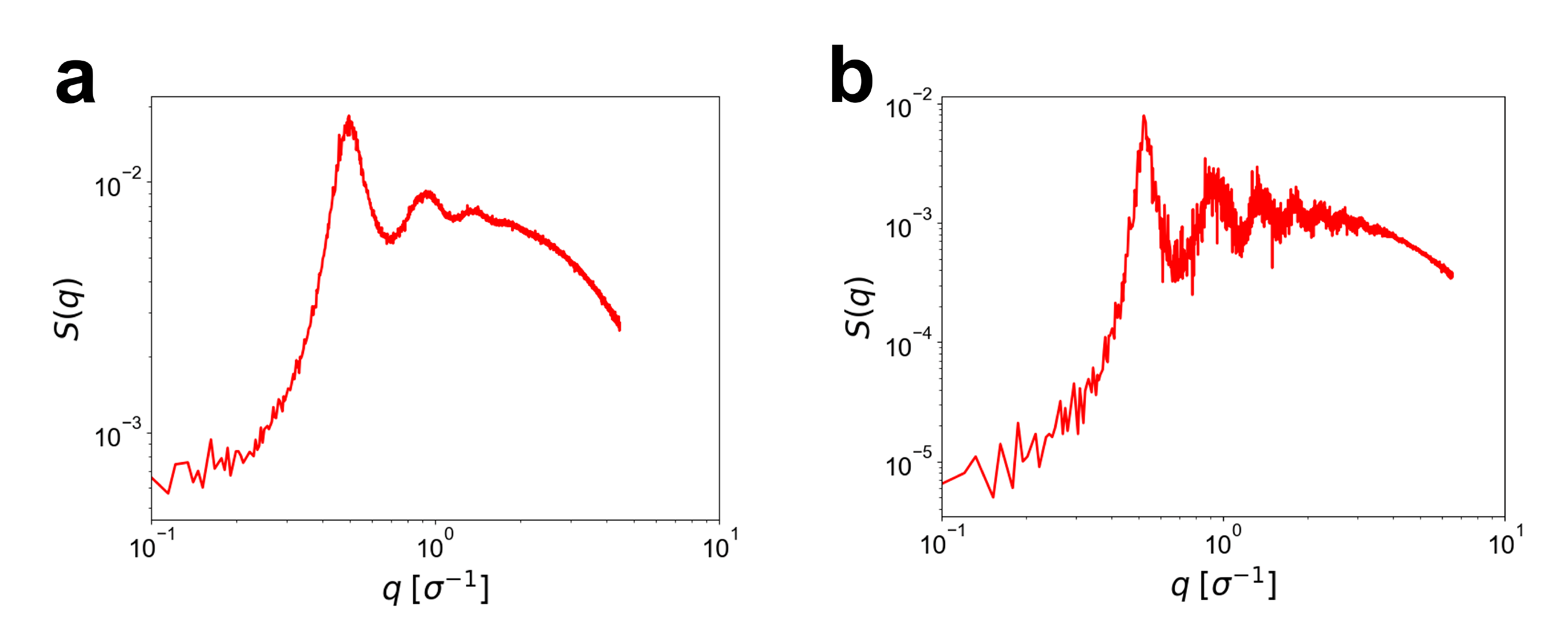}
\caption{The structure factors of the colloids for a) the disordered state ($Q = 24e$) and b) the crystalline state of the hybrid coacervate ($Q=150e$). The simulation parameters are $N = 120$ and $N=750$ for $Q = 24e$ and $Q=150e$, respectively; the values of $f = 0.2$ and $l_{B} = \sigma$ are the same for both systems. The structure factors here are calculated on the systems eight times larger than the systems shown in Figure 4 in the main text.}
\label{fig:d-SI}
\end{figure}

\bibliography{bibliography.bib}